\documentstyle[10pt,aaspp4,flushrt]{article}
\makeatletter

\@ifundefined{chapter}{\def\thebibliography#1{\section*{References\@mkboth
  {REFERENCES}{REFERENCES}}\list
  {\relax}{\setlength{\labelsep}{0em}
        \setlength{\itemindent}{-\bibhang}
        \setlength{\itemsep}{0pt}
        \setlength{\parsep}{0pt}
        \setlength{\leftmargin}{\bibhang}}
    \def\newblock{\hskip .11em plus .33em minus .07em}
    \sloppy\clubpenalty4000\widowpenalty4000
    \sfcode`\.=1000\relax}}%
{\def\thebibliography#1{\chapter*{Bibliography\@mkboth
  {BIBLIOGRAPHY}{BIBLIOGRAPHY}}\list
  {\relax}{\setlength{\labelsep}{0em}
        \setlength{\itemindent}{-\bibhang}
        \setlength{\itemsep}{0pt}
        \setlength{\parsep}{0pt}
        \setlength{\leftmargin}{\bibhang}}
    \def\newblock{\hskip .11em plus .33em minus .07em}
    \sloppy\clubpenalty4000\widowpenalty4000
    \sfcode`\.=1000\relax}}

\newlength{\bibhang}
\let\@internalcite\cite
\def\cite{\let\@citeleft(\let\@citeright)%
    \@ifstar{\citeyear}{\citefull}}
\def\citenp{\let\@citeleft\relax\let\@citeright\relax
    \@ifstar{\citeyear}{\citefull}}
\def\citefull{\def\astroncite##1##2{##1~##2}\@internalcite}
\def\citeyear{\def\astroncite##1##2{##2}\@internalcite}

\def\@citex[#1]#2{\if@filesw\immediate\write\@auxout{\string\citation{#2}}\fi
  \def\@citea{}\@cite{\@for\@citeb:=#2\do
    {\@citea\def\@citea{; }\@ifundefined
       {b@\@citeb}{{\bf ?}\@warning
       {Citation `\@citeb' on page \thepage \space undefined}}%
{\csname b@\@citeb\endcsname}}}{#1}}

\def\@cite#1#2{\@citeleft#1\if@tempswa , #2\fi\@citeright}
\def\@biblabel#1{}

\makeatother

\slugcomment{}

\newlength{\parindnt}
\newcommand{\PSbox}[3]{\mbox{\rule{0in}{#3}\includegraphics{#1}\hspace{#2}}}
\newcommand{\FigNum}[1]{\unitlength 1pt \begin{picture}(55,10)(-400,35) 
                        \put(0,0){Figure #1}
                        \end{picture}}

\newcommand\sig{$\sigma$}

\newcommand\msun{M$_\odot$}             
\newcommand\mdot{$\dot M$}

\renewcommand\approx{\mbox{$\sim$}}

\newcommand\approxgt{\mbox{$^{>}\hspace{-0.24cm}_{\sim}$}}
\newcommand\approxlt{\mbox{$^{<}\hspace{-0.24cm}_{\sim}$}}

\newcommand\ie{{\it i.e.}}
\newcommand\eg{{\it e.g.}}

\newcommand\etal{et~al.$\!$}

\def\Ref #1 {\lbrack {#1}\rbrack}
\def\vol#1  {{{\bf #1}{\rm,}\ }}
\def\etal   {{et~al.}\ }
\def\apj    {{ApJ{\rm,}\ }}
\def\apjl   {{ApJLett{\rm,}\ }}

\def\apjs   {{ApJS{\rm,}\ }}

\def\aa     {{A\&A{\rm,}\ }}

\def\pasj   {{PASJ{\rm,}\ }}

\def\gx339   {{GX 339-4}}
\def\g1124   {{GS 1124-68}}
\def\g2000   {{GS 2000+25}}
\def\g2023   {{GS 2023+33}}

\def\lmcx3   {{LMC X-3}}
\def\lmcx1   {{LMC X-1}}
\def\u1354   {{1354-64}}
\def\u1630   {{1630-47}}
\def\u1826   {{1826-24}}

\def\bhc     {{black hole candidate}}

\def\hr      {{Hard Ratio}}
\def\lr      {{Wide Ratio}}
\def\ler     {{Low Energy Range}}
\def\mer     {{Medium Energy Range}}
\def\her     {{High Energy Range}}

\def\ir      {{Intensity Range}}

\def\percm   {{cm$^{-2}$}}
\def\bhclfn   {{BHCLFN}}

\def\ccr      {{count-rate}}
\def\chisqrnu {{$\chi^2_\nu$}}
\newcommand{\fixedval}[1]{\multicolumn{#1}{l}{Values in parenthesis
      are fixed}}
\newcommand{\upperlim}[1]{\multicolumn{#1}{l}{Upper-limits are 3 \sig}}
\newcommand{\freqrange}[4]{\multicolumn{#1}{l}{{#4} Fits applied to
      values in the #2 -- #3 Hz range} }
\def\lac      {{LAC}}

\def\medf     {{Medium-Frequency}}

\def\highf    {{High-Frequency}}
\def\panel #1 {(panel {\it {#1}})}
\def\percm    {{cm$^{-2}$}}

\def\ppm      {{$\pm$}}

\def\rmssqrhz {{rms$^2$/Hz}}
\def\tensources {{1354-64, 1826-24, 1630-47, LMC~X-1, LMC~X-3,
    GS~2000+25, GS~2023+33, GS~1124-68, Cyg~X-1, and GX~339-4}}

\def\ginga    {{\it Ginga}}

\newcommand{\ud}[2]{$^{+ #1}_{- #2}$}

\def\t90{{T$_{90}$}}

\def\percm{{cm$^{-2}$}}

\begin{document}

\title{Quasi-Periodic Oscillations in Black Hole Candidates as an
Indicator of Transition Between Low and High States}

\author{Robert E. Rutledge$^{1,6}$, 
Walter H. G. Lewin$^2$, Michiel van der Klis$^{1,3}$,
Jan van Paradijs$^{3,4}$, Tadayasu Dotani$^5$, Brian Vaughan$^6$, 
Tomaso Belloni$^3$, Tim Oosterbroek$^{7}$, and Chryssa Kouveliotou$^{8}$}

{\small
\noindent $^1$ Department of Astronomy, University of California,
Berkeley, CA 94720\\
\noindent $^2$ 37-627, Dept. of Physics, Massachusetts Institute of
Technology, Cambridge, MA 02139\\
\noindent $^3$ Astronomical Institute ``Anton Pannekoek'', University of
Amsterdam, Center for High Energy Astrophysics, Kruislaan 403, 1098 SJ
Amsterdam, The Netherlands\\
\noindent $^4$ Physics Department, University of Alabama in Huntsville,
Huntsville AL 35899\\
\noindent $^5$ Institute of Space and Astronautical Science, 3-1-1 Yoshinodai,
Sagamihara Kanagawa 229-8510, Japan\\
\noindent $^6$ Space Radiation Laboratory, MS 220-47, California Institute of
Technology, Pasadena, CA 91125  \\
\noindent $^7$  Astrophysics Division, Space Science Department of ESA, 
ESTEC, 
P.O. Box 299, 
2200 AG Noordwijk, 
The Netherlands\\
\noindent $^{8}$ Universities Space Research Association, Huntsville, AL 35800 \\
}

\begin{abstract}
By comparing positions on a spectral color-color diagram from 10 black
hole candidates (BHCs) observed with \ginga\ (\tensources) with the
observed broad-band noise (0.001-64 Hz; BBN) and quasi-periodic
oscillation (QPO) variability, we find that the so-called ``Very High
State'' is spectrally intermediate to the Soft/High-State and
Hard/Low-State.  We find a transition point in spectral hardness where
the dependence of the BHC QPO centroid frequency (of GS~1124-68 and
GX~339-4) on spectral hardness switches from a correlation to an
anti-correlation; where the BBN variability switches from High-State
to Low-State; and where the spectral hardness of the QPO relative to
that of the BBN variability is a maximum.  This coincidence of
changing behavior in both the QPO and the broad-band variability leads
us to hypothesize that the QPO is due to interaction between the
physical components which dominate the behaviors of BHCs when they
occupy the Hard/Low and Soft/High States. We conclude that these QPO
should be observed from BHCs during transition between these two
states. Comparison with QPO and BBN behavior observed during the 1996
transition of Cyg~X-1 supports this hypothesis.

We also report 1-3 Hz QPO observed in GS~2000+25 and Cyg~X-1 in the
Hard/Low State, and we compare these to the QPO observed in 
GS~1124-68 and GX~339-4. 

\end{abstract}


\section{Introduction}

The X-ray (2-20 keV) phenomenology of galactic black hole candidates
(BHCs) has been widely studied (see \citenp{tanakalewin95} and
\citenp{mvdk95} for recent reviews).  The BHCs have many similarities
in their phenomenology, in that similar X-ray spectral hardnesses are
accompanied by similar fast-timing behavior (0.01-100 Hz).  Often, it
is this similarity of behavior alone which identifies an individual X-ray
source as a BHC, when this source has no measured mass-function, and
it presents correlated X-ray spectral and timing behavior which is
similar to that observed from BHCs which do have a (high, \approxgt 3
\msun) measured mass-function.

This X-ray phenomenology is largely composed of three ``states'',
historically tied to the observed source intensity, which has evolved
to be tied to the source photon spectral hardness.  The ``Soft/High''
state (HS; historically, when the 1-10 keV source flux was high) is
marked by a soft photon spectrum dominated by an ``ultra-soft''
black-body component, with $kT$ \approxlt 1 kev, and weak ($\sim$ few
\%) rms (root-mean-square) variability which is a power-law as a
function of frequency, of slope $\sim$ 1.  The ``Hard/Low'' state (LS)
is marked by a hard photon spectrum dominated by a power-law
component, and constant \%rms below some frequency ($\sim$ 0.01-10
Hz), decreasing roughly as a power-law above this frequency.  The
``Very High'' state (VHS) is marked by the presence of quasi-periodic
oscillations (QPO) with frequencies between 3-10 Hz.  In addition, an
``Off'' state (in which the source is quiescent) is generally included
in any phenomenology.  More recently, an ``Intermediate State'' (IS;
\citenp{belloni96b,belloni97,mendez97}) has been recognized, in which
the variability is band-limited, constant below some break frequency
($\sim0.1-10$ Hz) and decreasing above this, while the energy spectrum
shows both an ultra-soft black-body component and a power-law
component $>$10 keV.  The distinction made between this state and the
HS and LS are the simultaneous presence of the ultra-soft and
power-law energy spectral components, with band-limited broad-band
noise (BBN) variability and (in GS 1124-68) the presence of QPO at
source fluxes below that of the so-called ``High State''.

As a structure for phenomenologically characterizing BHCs, the state
paradigm lacks quantitative value.  For example, Cyg X-1 may be
observed in the Soft/High State.  One might well ask, how high?  Since
``high'' describes the observed flux, the uncertainty in distances to
BHCs makes it difficult to inter-compare the behavior of objects on
this basis. There currently exists no gradation of states, no formal
quantitative test to determine if a particular source occupies a state
or not, or ``how much'' it occupies a state.  The description of
additional ``states'', without somehow tying them through some
invariant observational parameter to other ``states'' does not
illuminate the relationship between the various states.  In addition,
it seems reasonable that the BHCs do not occupy ``states'' at all, but
that BHC behavior changes slowly as a function of its observables,
since individual BHCs are observed to have a wide range of
luminosities, spectral hardnesses and power density spectra, rather
than three or four of these.  As such, investigating BHC behavior as a
function of continual variables may be a useful observational
approach.

In an individual source, observational states are defined on the basis of
qualitative changes in its power-spectrum and energy-spectrum, which are
observed to be strongly correlated.  It seems likely that the
instantaneous accretion rate ($\dot{M}$) is the underlying variable
parameter responsible for the observed changes in source behavior.
However, between different sources, other parameters -- such as the
mass of the black hole, or the stellar type of its companion -- may also
come into play in determining the observed source intensity behavior.
Comparisons in behaviors between different objects therefore may
eventually provide  observational handles on these parameters.

The combined use of color-color diagrams (CCDs) and the
contemporaneous timing behavior -- usually indicated by the strength
of source variability and its functional dependence on frequency in
Power Density Spectra (PDS) -- has quite successfully revealed
regularities in the complex QPO and BBN timing behavior of GS~1124-68
\cite{takizawa97} and GX~339-4 \cite{miyamoto91}.  Following this
work, it was found that in the VHS, in which the QPO are found, the
PDS (apart from the QPO) can be like those in the Hard/Low State, or
like those in the Soft/High \cite{miyamoto93,takizawa97}. In a similar
vein, it has been shown that, as the fraction of the total intensity
due to the power-law spectral component increases above 10\%, the
\%rms variability increases, as well \cite{miyamoto94}; the
variability behavior of GS~1124-68 was demonstrated to be correlated
to the strength of its hard-spectral-component, which had been long
suspected.

The CCD vs. fast-timing s approach was first used in X-ray astronomy
to untangle correlated timing and spectral/intensity behavior of
non-pulsing neutron-star Low-Mass X-ray Binaries (LMXBs;
\citenp{hvdk89}).  Consequently,  it has become clearer what the
accretion rate is when the various distinct behaviors of these objects
are observed.  More detailed theory on the spectral and timing
behavior has been produced, as well as proposals for the relationship
between the behavior of neutron-star LMXBs and that of BHCs and
pulsars (\citenp{mvdk94a},b \nocite{mvdk94b}; see \citenp{mvdk95} for
review).

We have completed a study of the simultaneous spectral and timing
behavior of ten BHCs observed with \ginga\ (\tensources).  A
description of BBN behavior of these sources is being presented
elsewhere (\citenp{bhclfn98}; hereafter \bhclfn).  Here, we describe
and discuss the spectral and energy dependence of QPO in the four
sources which exhibit them in the \ginga\ data (GS~1124-68, GX~339-4,
Cyg~X-1 and GS~2000+25) and the strength and frequency dependence of
the underlying BBN.  These results are a summary of analyses presented
in greater detail elsewhere \cite{rutledgethesis}.

We have examined the source spectral state, as characterized by its
location on a CCD, and simultaneous timing properties.  We find that
BHCs show QPO typically when they have a spectral hardness ratio
between the Hard/Low State and Soft/High State.  When we follow the
timing behavior of the two objects which exhibit this QPO (GX~339-4
and GS~1124-68) as a function of spectral hardness, changes in
behavior of the QPO and the underlying BBN indicate that the so-called
VHS is a transition between the Hard/Low State and the Soft/High
State.  This has important implications for the origin of the QPO, as
an indicator of this transition, perhaps being produced by interaction
between two distinct physical processes responsible for the HS and LS
phenomena.

An important note on terminology: Historically, ``Low State'' was applied to
describe the behavior of BHCs when the source intensity was low, and
``High State'' for when it was high.  We are faced with a descriptive
problem, in that the behavior of the BHCs we describe herein -- aside
from intensity, and the presence of QPO -- is indistinguishable from
that observed from the ``Low State'', or the ``High State''.  It is a
common aspect of theories of BHC accretion to ascribe the Low State to
one physical component, such as a geometrically thick accretion disk
or a spherical corona, and the High State behavior to another, such as
a geometrically thin accretion disk (\eg\
\citenp{narayan94,abramowicz95,narayan95,chen95,chen96,narayan96,esin97}).
Therefore, throughout this paper we use the terms ``Low State'' and
``High State'' not to indicate source intensity, but behavior which is
ascribed to the respective physical components we mention above.


\section{The X-ray Data and Analyses}
\label{chap:bhcdata}

All X-ray data on \bhc s (BHCs) in the present work were obtained with
the Large Area Counter (\lac) on the the \ginga\ satellite
\cite{turner89}.  The \lac\ on-board \ginga\ had a geometric area of
$\sim$ 4000 cm$^2$, a nominal energy range of 1-37 keV, divided into
48 energy channels, with energy resolution of 18\% $(E/ 6 {\rm
keV})^{1/2}$. The time resolution of the data, depending on the
data-acquisition mode, was typically between 0.001-16 sec.

\subsection{Color-Color Diagram}
\label{sec:specanal}

The data used in the CCD are described in detail elsewhere (\bhclfn;
App. B in \citenp{rutledgethesis}).  Briefly, our data are drawn from
all data acquired by \ginga\ during observations of each of ten X-ray
black hole candidates (\tensources), which had energy resolution
sufficient to produce our hardness ratios (see also below).  All data
were corrected for background, deadtime, and aspect.  Following these
corrections, we produced hardness ratios using three bands -- the
\ler\ (2.3--4.6 keV), the \mer\ (4.6--9.2 keV) and the \her\
(9.2--18.4 keV).  These bands were combined into the \hr\ (\her\ /
\mer ) and the \lr\ (\her\ / \ler ), which we use to produce the CCD.
Each point represents 128 seconds of contiguous integration time.
Only points which have error bars in both colors showing significantly
measured hardness ratios ($>$4\sig) are shown (the remainder are
discarded).  Due to inadequate background subtraction, not all data
acquired by \ginga\ on a particular source is used in this CCD.  Thus,
the CCD is biased against fainter intensities, for which either the
background subtraction techniques employed or counting statistics in a
128-sec integration were inadequate.  After all selections were
applied, a total of 680 ksec (of 1500 ksec) of data from all sources
were used in the CCD.  The CCD is presented in Sec.~\ref{sec:CCD} .

In the lowest energy bin, the effect of galactic absorption is
minimized (although not eliminated) by using a high lower energy bound
(2.3 keV).  All BHCs in the present study have column densities
between 1.4 $10^{21}$ \percm and 6.0 $10^{21}$ \percm, except GS
2000+25 (8.4 $10^{21}$ \percm ) and 1630-47 (25 $10^{21}$ \percm;
\citenp{jan95}).  The effect of galactic absorption in the 2.3-4.6 keV
energy range is small -- for a black-body of kT=0.5 keV, the \lr\
changes by $<$10\% between 1.4 $10^{21}$ \percm and 6.0 $10^{21}$
\percm; at 25 $10^{21}$ \percm , this becomes a 30\% effect in
the 2.3--4.6 keV energy range.
								     
Throughout the present work, we refer to spectral ``states'',
typically as indicated by the contemporaneous \lr\ value.  For
brevity, we use a progression of letters (A, B, C, D..., J) which
indicate a range of \lr\ space, each half a decade wide in logarithmic
space, increasing in sequence with increasingly hard spectrum (A is
$-4.0 \leq$\lr\ $\leq -3.5$, B is $-3.5 \leq$\lr\ $\leq -3.0$, etc.).
The width of these ranges was arbitrarily chosen, with the idea that
the source behavior does not vary significantly within a single
range. This is not always the case, particularly when QPO are
observed, and it then becomes necessary to further subdivide the data
based on the spectral values into ``sub-states'' (\eg\ F1, F2, ...).
However, the letters A-J, are retained for convenience in referencing
the timing analysis to a simultaneous spectral indicator.

\subsection{Timing Analysis}

\label{sec:bhcpds}

In Table~\ref{tab:timingdata}, we show the quantity of data which are
available to us for the timing study, separated by source and time
resolution.  In all, 853 ksec are available for the fast-timing
analysis; of this, 305 ksec have time resolution of or better than
0.06250 sec.

An extensive description of our PDS production, including mathematical
definitions, normalization, and computation of correction factors for
dead time and background, is to be found elsewhere
\cite{rutledgethesis}.  We repeat the pertinent details here.

\subsubsection{Composite PDS}

To make efficient use of all available data, each observation
was used to produce any (or all) of three possible types of
FFTs, which we define here:
\begin{itemize}
\item {\bf Low Frequency FFT}: Data is binned to 16.0~s, with a total
  duration of 1024~s, using 64 bins of data.  The resulting
  frequency range is 0.0009765--0.03125 Hz. 
\item {\bf Medium Frequency FFT}: Data is binned to 0.0625~s, with a
  total duration of 64~s, using 1024 bins of data.  The resulting
  frequency range is 0.015625--8 Hz. 
\item {\bf High Frequency FFT}: Data is binned to 7.8125 msec, with a
  total duration of 8~s, using 1024 bins of data.  The resulting
  frequency range is 0.125--64 Hz. 
\end{itemize}

Thus, a single observation with a duration of 1500 s and a time
resolution of 7.8 msec is used to produce 187 High-Frequency FFTs, 23
Medium-Frequency FFTs and 1 Low-Frequency FFT.

Using raw data selected from those listed in
Table~\ref{tab:timingdata} (listed in greater detail in App.~B of
Rutledge \citenp*{rutledgethesis}), we produced FFTs in three separate
energy bands: the \ler\ (=2.3--4.6 keV), \mer\ (=4.6--9.2 keV) , \her\
(=9.2--18.4 keV) and in the summed \ir\ (=2.3--18.4 keV).  This
produced a library of Low-, Medium-, and High-Frequency FFTs in three
independent photon energy ranges and one summed photon energy range
from which we draw to produce PDS, based on the selection criteria we
apply for any given analysis.

To produce a Composite PDS, individual FFTs are selected by the \lr ,
measured during the time period covered by the FFT.  Each individual
FFT is made into a single PDS, its calculated dead-time corrected
Poisson Level subtracted, and the resulting power corrected for
background and instrumental deadtime.  Finally, all the resulting PDS
are averaged together into a single PDS.  The uncertainty in the power
at each frequency is taken to be the measured standard deviation in
the power divided by the square root of the number of measurements.
Thus, the uncertainty in the power contains any (non-biased)
systematic uncertainty in the Poisson-subtraction, dead time and
background \%rms correction factors.

These PDS are logarithmically rebinned, and overlapping frequency
ranges are dropped in the PDS with the lesser integrated
time.  For example: with 50 Medium-Frequency FFTs and 200 High
Frequency FFTs, the Medium-Frequency FFTs represent 64 s $\times$ 50
= 3200 s  of data, while the \highf\ FFTs represent 8 s $\times$
200 = 1600 s of data.  Therefore, all the data $<$8 Hz would be
dropped in the High-Frequency PDS, and then the two PDS are merged.

The final frequency range of the composite PDS is 0.00097--64 Hz or
less if data of sufficient contiguous duration or time resolution are
not available, and is normalized as \rmssqrhz , and corrected for dead
time and background.

The data of the high, medium and low frequency range parts of the
single PDS are not necessarily taken at the same time.  Thus, this
type of PDS is subject to systematic uncertainties due to changes in
the PDS within a single selection criteria.  In addition, there may be
more data available in one frequency range than in others.

For each of the ten individual objects, we produced four composite PDS
(of the Low, Medium, and High energy Range, and for the full Intensity
Range) from data selected by \lr\ to be in each of the coarse spectral
hardness bins (A, B, C, ...), as illustrated in
Fig.~\ref{fig:colorccd} and described in Sec.~\ref{sec:CCD}.  We used
these PDS to investigate the dependence of BBN variability on spectral
hardness, to search for the presence of QPO and understand the
dependence of QPO properties on spectral hardness and photon energy.

There are, at times, intensity ``spikes'' which occur in the data, a
few msec in duration, which are typically not due to source
variability, but instead to a variety of background sources (some of
these spikes were due to geomagnetically trapped  particles
from orbiting nuclear power generators.  These events, and the
observed 2-37 keV PHA spectra, are discussed further elsewhere
\cite{rutledgethesis}).  We systematically excluded these spikes
from our analysis, as they can produce significant, measurable power
across all frequencies.  When a single time bin is found in which the
number of counts is 20 \sig\ (calculated from the intrinsic scatter of
the data used for that particular FFT) greater than the mean number of
counts, that value was replaced by a value randomly selected from a
set of bins in the data used for that FFT.  Such events, even if real,
are identifiable as discrete events, which are not our primary focus.

\subsection{Parameterized Fits to the PDS}

In practice, the models adopted to fit a PDS may not be ideal, or even
formally acceptable according to statistical tests.  Discrepancies may
be systematic -- for example, due to the skew of a quasi-Lorentzian
peak which is not accounted for in the model -- or they may be due to
excursions from simply parameterizable continuum behavior in the PDS
-- that is, deviations from the model which are non-biased, like
random scatter, but statistically significant.  There exists no {\it a
priori} reason that such biased and noisy PDS should not exist.
Thus, while it is desirable to fit the models as closely to the data
as possible, there can be a trade-off between the accuracy and
simplicity of a model.  These parameterizations are not physically
motivated, and thus their utility is to accurately describe PDS using
a small set of numbers.

For the present study, we desire simple parameterizations which
adequately describe PDS from all 10 sources.  We avoid the
introduction of new parameters, apart from switching to a more
reasonable parameterization.  We use a maximum-likelihood fitting
method (least $\chi^2$; \citenp{press}), and quote the 1$\sigma$
single parameter uncertainties, the reduced $\chi^2$, and the number
of degrees of freedom.

We use a number of models, most of which have become widely used in
describing PDS of X-ray binaries, to fit the observed power (P) as a
function of frequency ($\nu$) to the PDS.
\label{sec:themodels} The model shapes we used to fit to the real
source variability are all continuous functions of frequency, and
include:

\begin{itemize}
\item Power-Law, P$\propto \nu^{-\alpha_1}$
\item Flat-Top Single-Power-Law, a constant power below some frequency
  $\nu_1$ breaking to a Power-Law with slope $\alpha_1$ above this
  frequency. 
\item Broken Power-Law, represented by one power-law slope
  ($\alpha_1$) below a critical frequency $\nu_1$ and a different,
  typically steeper, power-law with slope $\alpha_2$.  If we find $\alpha_1$
  \approxlt 0.3 (that is, if the first power-law is very flat), then
  we call this model ``Flat-Top Single-Power-Law'' instead of a
  ``Broken Power-Law''.  This is a definition of convenience, having
  no theoretical or observational impetus.
\item Flat-Top Broken-Power-Law,  a constant power below some frequency
  $\nu_1$ breaking to a Power-Law with slope $\alpha_1$ above this
  frequency, breaking again to a (typically steeper) Power-Law
  $\alpha_2$ above frequency $\nu_2$, with $\nu_1<\nu_2$. 
\item Double Flat-Top Single-Power-Law,  a model in which the power is
  constant below some frequency $\nu_1$, breaking to a power-law of
  slope $\alpha_1$, which evens out into a constant value again at
  some higher frequency $\nu_2$($> \nu_1$), and then breaks again into
  a power-law of the same slope ($\alpha_1$) at some higher frequency
  $\nu_3$.  
\item Flat-Top Power-Law Exponential, a model in which the power is
  constant below some frequency $\nu_1$, and continuous with the
  product of a power-law and an exponential function above this
  frequency. 
\end{itemize}
 
In fitting models to the data, we employ one of the above models,
which we call the continuum model, or Broad-Band Noise (BBN) model,
and a model to account for quasi-periodic oscillations from:

\begin{itemize}
\item Lorentzian, used to model quasi-periodic oscillations, which
  has a centroid frequency $\nu_c$ and Full-Width at Half-Maximum
  (FWHM; $\Gamma$), and a normalization related to the \%rms variability.  
\item Harmonics, sometimes added to account for the appearance of
  Harmonics and sub-Harmonics to the QPO; the frequencies and FWHM
  values of these Harmonics are constrained to be 0.25, 0.5, 2.0 and
  4.0 that of the Fundamental, except where noted. 
\end{itemize}

The \%rms variability  of the Lorentzians and Harmonics are integrated
between $-\infty $ and $\infty$ Hz.

\subsection{Energy Dependency of the PDS }

We present the results of the fits to the four sources in which we
observed QPO (GS~1124-68, GX~339-4, Cyg~X-1, and GS~2000+25) in four
energy ranges.  For each PDS, we indicate the mean source spectrum by
the \lr\ (=\her/\mer).  Finally, we fit the PDS with BBN and QPO models
to determine the parameterization of the PDS and the \%rms variability
in each spectral state.


\section{Results}

\subsection{Color-Color Diagram}

In Fig.~\ref{fig:colorccd}, \label{sec:CCD} we show a color-color
diagram (CCD) of all X-ray data in the present study, showing the \lr\
vs. the \hr .

The dashed-line divisions on the figure are artificial boundaries,
separating the spectral parameter space into equally sized areas based
on the \lr .
\label{sec:alphabet} These divisions are labelled with the alphabetic
shorthand (A, B, C, D, etc.), encompassing two such ranges,
corresponding to a decade in magnitude of the \lr .

The approximate locations of each source are labelled on the CCD.  The
CCD shows that, for most of these sources, spectral evolution in these
colors produces roughly parallel tracts which run diagonally across
the CCD.  These tracks show smooth variation in the spectrum within
the CCD, although there are discontinuities -- likely due in
part to short observing times relative to the time-scale of spectral
evolution.

We have approximately indicated the positions on the CCD where BBN
behavior expected from the various ``states'' was observed in the PDS
of sources in that area.  Sources which are spectrally soft, have PDS
which have a single-power-law dependence on frequency, and present
only weak ($\sim$few \%rms) variability (1-20 keV) occupy the area
labelled ``High State''.  Sources which are spectrally hard, have
complex PDS -- typically, but not always, constant below a frequency
$\sim$0.1-10 Hz, and decreasing as a power-law at higher frequencies
-- and have strong ($\sim$20-50\%rms) variability (1-20 keV) occupy
the area labelled ``Low State''.  The behavior of High State and Low
State variability will be presented in detail elsewhere (\bhclfn).
Sources which present QPO occupy the spectral range labelled ``Very High
State''.

Some sources were only observed to occupy the Low State (such as
GS~2023+33, 1630-47, 1826-24, and Cyg~X-1).  Others were observed only
in the High State (1354-64 and LMC X-3).  Some were observed to
traverse two or more of these states (GS~1124-68, GX~339-4, and
LMC~X-1 -- although the observation of LMC X-1 in the spectral region
corresponding to the Very High State did not have sufficient time
resolution to confirm the presence of QPO).

In Fig.~\ref{fig:qpopresentbig}, we show on the same scale as
Fig.~\ref{fig:colorccd} the points at which QPO were identified (by
eye) as present in the PDS of several sources in the present study.
In addition, we include a point from QPO from GX~339-4 in a previous
study \citenp{grebenev91}, observed with the ART-P spectrometer, with
a centroid frequency of 0.8 Hz, FWHM of 0.15 Hz, and a \%rms
variability of 7.0\ppm 0.8\% in the 3-25 keV energy range.  When this
QPO was detected from GX~339-4, the energy spectrum was power-law,
with a photon slope of 1.733\ppm 0.004.  This places GX~339-4 at (\lr,
\hr) vector of (0.32, 0.40), found by propogating the observed photon
spectral model through the \ginga\ response (see
Sec.~\ref{sec:spect}).  While the \lr\ of the sources in the present
study span four orders of magnitude, the QPO is largely clustered
within $<$1 order of magnitude, located close to a single locus in the
CCD. Sources which were observed to exhibit QPO were also observed at
other times at different positions on the CCD, but with no QPO.

Fig.~\ref{fig:hrsources} shows the \lr\ for each of the ten sources
(same data as Fig.~\ref{fig:colorccd}, described above).  There are
three sources -- GS~2000+25, LMC~X-1, and LMC~X-3 -- in which some
observed \lr\ values are similar to those observed from GX~339-4 and
GS~1124-68 when 3-10 Hz QPO were observed (the Very High State
spectral range).  Timing analysis of all observations (but one) of
GS~2000+25 produced upper-limits to QPO strength well above that
observed from GX~339-4 and GS~1124-68; in the one observation near
\lr=0.1 (at the edge of the Very High State spectral range), 2.6 Hz
QPO from GS~2000+25 was observed (see Sec.~\ref{sec:g2000lsqpo}).
LMC~X-1 and LMC~X-3 were observed with time-resolution of 16 sec --
too long to measure QPO at 3-10 Hz.  Thus, we do not observe any
BHCs which exhibit spectral hardness in the Very High State
spectral range, but no QPO (with upper limits below the observed
strengths in GX~339-4 and GS~1124-68).

\subsubsection{Intrinsic Source Spectrum to \lr\ Conversion}
\label{sec:spect}
As the \lr\ (as measured by \ginga ) is the parameter to which we tie
the observed behavior, we provide a figure to relate this ratio to the
observed photon spectral parameters of a power-law, black-body, and
the relative strength of the two.  To do so, we simulated photon
energy spectra with black-body temperatures of $kT=$0.1, 0.25, 0.4,
0.55, 0.7, 0.85, 1.0, and 1.4 keV; with photon power-law spectra of
slopes = 0.9, 1.3, 1.7. 2.1, 2.5, and 2.9.  (We use the black-body
spectrum as a simple parameterization, although more physically
reasonable models, such as a disk black-body spectrum, may more
accurately describe the spectrum in this energy range; 
\citenp{mitsudadiskbb,makishima86}).  We combined these two spectral
components with different relative strengths, indicated by the ratio
($X$) of 2-20 keV photon flux (photons/sec) of the power-law to that
of the black-body, for values of X in the range of 0.0001-100.

The relationships between $X$ and \lr\ for these assumed photon
spectra are shown in Fig.~\ref{fig:speclr}.  To have a \lr\ \approxlt
0.01, a BHC must be dominated by the black-body (that is, ultra-soft)
component.  To obtain a \lr\ \approxgt 0.1, BHCs with photon spectral
slopes $<$ 2.2 must be dominated by the power-law photon spectral
component.  At steeper spectral slopes, and in between 0.01--0.1
where most of the QPO in GS~1124-68 and GX~339-4 are observed, the
\lr\ is a complex function of $X$, $kT$, and the photon spectral
power-law slope.


\subsection{Power Density Spectra}

Here, we present the parameters describing the QPO and their energy
dependence, including the strength of the simultaneously observed BB
variability and the models used to describe them. The detailed results
pertaining to BBN variability when QPO are not observed, and the
detailed parameters and energy dependence of the BBN variability when QPO
are observed are presented elsewhere (\bhclfn; \citenp{rutledgethesis}).

\subsubsection{Cyg X-1} 

Of the measured \lr\ values for Cyg X-1, 80\% were in the range
0.13--0.51, which is within the Low-State described in
Sec.~\ref{sec:specanal}.

When Cyg X-1 exhibited QPO, the values of its \lr\ are largely in one
of two ranges: 0.12--0.15 and 0.16--0.18. These in turn correspond to
two different observational periods (90/129--131 and 91/156--157,
respectively). We sub-divided these data into two spectral groups:
\lr=0.1--0.15 (G1), and 0.15--0.32 (G2).  The QPO parameters from the
best-fit models are given in
Table~\ref{tab:cygx1compositepdsqpoparams}.

In Fig.~\ref{fig:cygx1splitg}, we show the PDS of the G1 and G2
spectral ranges, with the power renormalized as $\nu P$. For all
energy ranges, the PDS are identical above 10 Hz.  Below 10 Hz, the
PDS are different, and appear to be complex functions of frequency.
Most evident is the presence of a QPO peak near 1 Hz in the PDS of the
G1 data.  The best fit BBN variability model, without a QPO Lorentzian
near $\sim$1 Hz, has a $\chi^2_\nu>13$, which decreases to $\sim$1.6
when the Lorentzian is added.  The \%rms variability of this QPO
decreases with increasing energy range from 11.7\ppm 0.5 in the \ler\
to 8.2\ppm 0.6 in the \her .

We also fit a Lorentzian to the G2-data PDS. In the \ir\ a peak was
found at 3.14 Hz, with FWHM 0.7\ppm 0.3 Hz.  These values were fixed
for the \ler , \mer , and \her\ PDS, in which the \%rms was consistent
with being colorless.

This QPO peak is very broad ($Q\sim$1), which would usually result in its
being called a ``noise'' feature, as QPO is a term traditionally
reserved for higher $Q$ signals.  We attempted to distinguish this
feature from the QPO of GX~339-4 and GS~1124-68 (see Sec.~\ref{sec:R})
based on the QPO spectral hardness as a function of the source photon
spectral hardness, but found we were unable to do so on this basis.
We therefore forward the hypothesis that this feature is related to
the QPO observed in the data from GS~1124-68 and GX~339-4.

\subsubsection{GX 339-4}

Of the measured \lr\ values for GX~339-4, 80\% were in the range
\lr=0.044--0.56. 

We closely examined the data obtained in the September 1988
observations of GX 339-4, divided by source spectrum, which is the only
period during which the \lr\ was in the F spectral range, and during
which the VHS QPO was observed.  In Fig.~\ref{fig:gx339qpohidpds}, we
show the \lr\ as a function of time during these observations.  The
source spectrum (measured by \lr) hardens between t=0 and 2 days; over
the next two days, the source spectrum globally softens, while
exhibiting short-term reversals -- variously hardening and softening.

We divided the data by spectrum, and labeled the spectral regions
(F1-4; all in the F-state) on the hardness-intensity diagram (HID) in
the upper-right panel of Fig.~\ref{fig:gx339qpohidpds}.  We recognize at
least three separate tracks in the HID -- a nearly constant \ccr\ 
track (near I=5800 c/s 2.3--18.4 keV, crossing the regions labeled
``F3'' and ``F4''), a nearly constant \lr\ track (near \lr=0.045 --
labeled as ``F2''), and a steeply sloped track (the \lr\ region
labeled ``F1'').  These tracks correspond to those found in Fig. 5 of
Miyamoto \etal \cite*{miyamoto91} and Fig. 5.28 of Ebisawa
\cite*{ebisawathesis}. 

We used the \medf\ and \highf\ FFTs in each spectral region to produce
Composite PDS, in all energy ranges.  We show these PDS in
Fig~\ref{fig:gx339qpohidpds}, in the bottom 16 panels.  The scale of
all PDS is given in the bottom left panel. When spectrally hardest (F3 and F4),
the source shows band-limited noise, breaking at a frequency of
\approx few Hz, in addition to QPO which is strongest in the 9.2-18.4
keV energy band.  When spectrally softer (F1 and F2), the PDS show
less variability, and a freqency dependence  which is more like a
power-law, or perhaps band-limited with a break-frequency near 0.02
Hz.

\label{sec:gx339vhsbb}

We fit the Composite PDS produced from the F1-4 data of GX~339-4.  The
models used were Single-Power-Law (for F1),
Double-Flat-Top-Single-Power-Law (for F2, with  the low-frequency
flat-top $\nu_1$ frozen to below 0.01 Hz), and
Flat-Top-Single-Power-Law (states F3 and F4).  For all these, we also
fit Harmonic QPO, allowing the \%rms normalization in the first two
harmonics and sub-harmonics to vary.  Typically, the \her\ PDS did not
constrain the models well; for these PDS, we froze the BBN model
parameters (except the \%rms) at their best fit values found from the
\ir\ PDS.

In Table~\ref{tab:gx339vhsrms}, we list the QPO and BBN \%rms values
found from these fits. In all PDS, the strength of QPO increases with
increasing photon energy.  The strength of the QPO does not show a
monotonic dependence on  spectral hardness of GX 339-4.  Looking at
the PDS, one might be misled to believe that the \%rms of the QPO in
the \ler\ PDS increases with increasing spectral hardness; our model
fits do not bear this out.  One is apparently misled by the relative
strength of the BBN power, which accounts for a dominant portion of the
\%rms at the QPO frequency when the spectrum is hard, but considerably
less when the spectrum is soft.  While we do see dependence between
the QPO \%rms in the \ler\, it is not monotonic, nor pronounced, with
the \%rms values in the range of 1.1--2.9\%.

\label{sec:gx339f1f4}

The QPO FWHM and centroid frequency exhibit some dependency (again,
non-monotonic) on spectral hardness. In the \ir\, the FWHM increases from
0.5, to 1.0, to 3.6, and back down to 2.3 as the spectrum hardens
between states F1--4.  The centroid frequency goes from 5.8, to 6.3, to
7.4, to 6.3 as the spectrum hardens between states F1--4.  In all
states, the QPO parameters (centroid frequency, FWHM) are energy
independent.

\subsubsection{GS 1124-68}
\label{sec:g1124pds}

Of the measured \lr\ values for GS~1124-68, 80\% were in the range
0.00098--0.38.  

\label{sec:takisec}

The QPO properties in the summed 1.2--18.4 keV range exhibited by
GS~1124-68 have been examined in detail \cite[hereafter,
T97]{takizawa97}.  We here re-examine these properties in our three
energy ranges, as well as the \ir, using the \lr\ spectral indicator. 

Over short periods of time ($\sim$ hours) there is typically very
little evolution in the source spectral hardness, less than 0.5 dex in
hardness ratio (see T97, \bhclfn).  We therefore produced PDS over
single observational epochs.  We adopted the same time-periods used by
T97, and the same labels for these time periods (by number 1-15), with
two exceptions: 1) we did not distinguish between observation 5L and
5H (where ``L'' stands for periods within the trough of a
``flip-flop'' -- see T97 for a description of this phenomenon), while
``H'' stands for the elevated \ccr\ periods around the flip-flop) --
they are joined under the single header ``5''; also, we split
observation 11 into 11a and 11b.  The time limits for the data used
here are listed in Table~\ref{tab:g1124takidata}, along with the
frequency range of the resultant PDS, and the mean \lr\ for each
observation period.  The uncertainty listed in the \lr\ is the 1\sig\
spread of observed 128-second values (the relatively small values
demonstrate that the amount of spectral evolution during a single
observational epoch is typically small).

As found previously (T97, and references therein; \citenp{belloni97}),
QPO is detected during observations 1, 2, 3, 5, 10, 11, and 15.  The
QPO during observation 11 appears is strongest in the second half
(11b).

In Table~\ref{tab:g1124takiparams}, we list the centroid frequency,
FWHM, and \%rms of the fundamental QPO in each PDS in which QPO was
identified by T97.  In addition, we find (weak) QPO in Observation \#
8 and 11a.  The fundamental was identified for each observation as
that QPO peak with the greatest \%rms in the 9.2--18.4 keV range, or,
if none are identified, in the 4.6--9.2 keV range.

The QPO was fit as a series of harmonics, with frequencies and FWHM
values which are factors of 0.25, 0.5, 1.0, 2.0, and 4.0 of those of
the fundamental.  Typically, only the sub-harmonic at 0.5 $\nu_c$ and
the second harmonic at 2 $\nu_c$ were invoked (permitting the best-fit
value of the \%rms in each to be determined).  For a few, however,
three or four harmonics have significant \%rms values.  These
harmonics were studied in detail elsewhere (T97;
\citenp{belloni97,rutledgethesis}).

In all cases, the \%rms of the fundamental increases with increasing
photon energy (except observation \#15, where the S/N is too low to
permit detection in the \her ). 

In addition to the detailed parametric fits to the PDS of the
individual observations, we also produced PDS using the F1-4 division
done for GX~339-4, for qualitative comparison with that source.  These
are shown in Fig.~\ref{fig:gs1124qpohidpds}, produced similarly as for
GX~339-4 (see Fig.~\ref{fig:gx339qpohidpds}).  When the source is
spectrally hard (F3 and F4), the PDS show band-limited noise with a
break frequency of a few Hz, along with QPO which is strongest in the
9.2-18.4 keV energy band.  When the source is spectrally soft (F1 and
F2) the PDS shows weaker variabilty, and a frequency dependence which
is either power-law (F2) or band-limited noise with a break frequency
of \approx 0.1 Hz.  This behavior is similar to that observed from
GX~339-4 using the same spectral hardness division. 

\subsubsection{GS 2000+25}

Of the measured \lr\ values for GS~2000+25, 80\% were in the range
0.021--0.23.  The values of the \hr\ are typically higher than those
observed for other objects with similar \lr\ values.

In a check of PDS of individual observation periods, a period of
significant variability was found on day 252 of 1988 (Sept 8,
$\sim$140 days into the outburst), between
15:45 and 17:24 UT.  During this period, the source was in a spectral
state with \lr=0.09--0.1.  We show the PDS, from an average of 44
Medium-Frequency FFTs, in Fig.~\ref{fig:g2000qpopds}.  We fit the data
with a Flat-Top-Power-Law function with an additional Lorentzian
component to account for QPO near 2 Hz.  We find that the QPO
component is required by the data in the \ler\ (compare \chisqrnu= 3.4
for 29 degrees of freedom, or dof,  without the QPO component to 0.7
for 26 dof with the QPO component).  We fit the Flat-Top-Single-Power-Law
plus QPO model to the \ler , \mer, and \her\ PDS made from data taken
during the period when the LS QPO was observed.  The best-fit model
parameters are listed in Table~\ref{tab:gs2000qpopdsparams}.  The
centroid frequency (\ir\ PDS) is 2.63\ppm 0.06, FWHM is 1.7\ppm 0.2,
and the \%rms variability is consistent with being independent of
energy.

\label{sec:g2000lsqpo}

We produced a PDS using data with a \lr\ consistent with
that observed from GS~1124-68 and GX~339-4 while they exhibit VHS QPO
(F range). Visual inspection of the PDS from this observation finds no
QPO peak.  After finding the best-fit BBN models in the \ir\ we added a
Lorentzian, with $\nu_c$=4, 5, and separately 6 Hz, with FWHM=1.0 Hz,
to obtain 3\sig\ upper-limits of the \%rms variability of such
component, which were 7.2\%, 4.8\%, and 7.5\% (respectively).  In the
\her\ the corresponding 3\sig\ upper-limits were 29\%, 26\%, and 30\%.
As these limits are greater than the measured strength of VHS QPO in
GX~339-4 and GS~1124-68, we cannot exclude the presence of VHS QPO in
this data.

\subsection{The Energy Dependence of the QPO}

In Fig.~\ref{fig:vhsqpo}, we show the centroid frequency and FWHM of
the QPO in GX~339-4 and GS~1124-68 as a function of spectral hardness.
In both GX~339-4 and GS~1124-68, the centroid frequency is dependent
on spectral hardness. The centroid frequency is greatest near
\lr=0.055 (\ppm 0.01 approximately), and decreases as the source
becomes spectrally harder and as the source becomes spectrally softer.
The same is true for the FWHM of the QPO -- it is greatest at
\lr=0.055, and smaller as the source is spectrally harder or softer.
We also include the QPO of Cyg~X-1 and GS~2000+25 in this figure,
which show roughly consistent behavior. 

In Fig.~\ref{fig:vhsbbqporms}, we show the \%rms variability as a
function of photon energy for QPO and BBN variability in GS~1124-68
and GX~339-4.  In both GS 1124-68 and GX 339-4 (panels a and d), the
VHS QPO \%rms (averaged over all observations) increases with
increasing energy.  The BBN average power (panels b and e), however,
increases from the \ler\ to the \mer ; it then remains constant or
decreases between the \mer\ and \her .

In GS~1124-68, when there is no QPO observed (panel c), the BBN
average power increases with increasing photon energy across all three
energy ranges. To produce this panel, we used observations 6, 7, 9,
12, and 14, in which VHS QPO were not observed, but the source
spectrum was similar to that during times when VHS QPO were observed
(in the range 0.01--0.03 vs. 0.013-0.155).  In the \ler\ and \mer\ the
average broad-band power is considerably larger when QPO is present
than when QPO is absent (compare 9.8\ppm 0.1 and 18.3\ppm 0.5 \%rms
with 2.8\ppm 0.05 and 10.0\ppm 0.2 \%rms).  However, in the \her , the
aveage power is comparable (15.4\ppm 0.8 \%rms with QPO, 17.1\ppm 0.4
\%rms without).  The average ratio of BBN variability in the \her\ vs
the \ler\ when QPO is absent is 6.1 \ppm 0.2 vs. only 1.6 \ppm 0.1
when QPO is present.  This may imply that the presence of QPO
supresses BBN variability in the high energy range.

In Fig.~\ref{fig:lsqpo} we show the energy dependence of the LS QPO in
Cyg~X-1 and GS~2000+25.  The QPO in these sources is slightly
anti-correlated with energy (in Cyg X-1) or constant (in GS~2000+25).
This is quite a different energy dependence than observed in the QPO of
GX~339-4 and GS~1124-68 (see Tables~\ref{tab:g1124takiparams} \&
\ref{tab:gx339vhsrms}, which are always spectrally hard.  In
particular, observations \#1, 2, and 3 of GS~1124-68, which have
spectral hardness similar to Cyg~X-1 and GS~2000+25 here, are not
spectrally flat or soft, but spectrally hard. This may imply a
different mechanism for these QPO of Cyg~X-1 and GS~2000+25 from those
of GS~1124-68 and GX~339-4.  (We are unable to characterize the LS QPO
of GX~339-4 reported previously by Grebenev \cite*{grebenev91}, as no
information on the energy dependence of the variability is available).

\subsection{The Variability Ratio}
\label{sec:R}

To compare the spectrum of the QPO relative to that of the broad-band
variability in the sources Cyg~X-1, GS~2000+25, GX~339-4, and
GS~1124-68, we define the variability ratio:
\label{sec:qpohard}
\begin{equation}
\label{eq:rr}
R_{QPO, BB} ={\frac{\rm \%rms_{QPO}(9.2-18.4 keV)}{\rm
     \%rms_{QPO}(2.3-4.6  keV)}} \div {\frac{\rm \%rms_{BB}(9.2-18.4 keV)}{\rm \%rms_{BB}(2.3-4.6  keV)}}
\end{equation}

\noindent which is a ratio of two ratios, each being the ratio of the
\her\ \%rms variability to the \ler\ \%rms variability, in the QPO and
(separately) the broad-band components.  Values of the variability
ratio which are consistent with 1.0 mean the QPO and BBN components
have the same energy spectrum; values which are greater than 1.0 mean
the QPO is spectrally harder than the BBN component.  The variability
ratio is independent of the source spectrum, in the sense that if the
spectrum changes while the \%rms variability as a
function of frequency remains the same, the variability ratio also
remains the same.
     
For GS~2000+25 and Cyg X-1, the spectrum of the QPO is the same as, or
slightly softer than the spectrum of the BBN variability.  The
variability ratio of GX~339-4 is significantly higher, which indicates
that the VHS QPO is spectrally harder than the BBN variability.  The
same is often true for GS~1124-68: the VHS QPO is spectrally harder
during observations 1, 3, 5, 10, and 11b.  During observations 2, 8,
and 15, however, the QPO spectral hardness is within a factor of 2
(spectrally harder) of the spectral hardness of the BBN variability.

\label{sec:RR} 
In Fig.~\ref{fig:rrvswr}, we show the variability ratio as a function
of the \lr , for the four objects from which we identify QPO.  For
GS~1124-68 and GX~339-4, the value of the ratio is high (\approxgt 4)
while the \lr\ is in the range 0.03-0.08. While GX~339-4 was not
observed with QPO outside of this range, GS~1124-68 was observed with
spectral hardness both above and below this range, where the ratio is
observed to be low (\approxlt 2).  For Cyg~X-1 and GS~2000+25, from
which QPO were observed in their LS, the value of the ratio was
observed to be low (\approxlt 2).  The similarity in the value of the
ratio observed from GS~1124-68, Cyg~X-1, and GS~2000+25 while they
have similar spectral hardness supports the argument that the QPO of
these objects are of similar origin.  The QPO centroid frequencies of
GS~1124-68 were 3-5 Hz (Observations \#1-3;
Tables~\ref{tab:g1124takidata} and \ref{tab:g1124takiparams}), which
are low, but not identical to the $\sim 1$ Hz frequencies of Cyg~X-1
and GS~2000+25.  Fig.~\ref{fig:rrvsvc} illustrates the correlation
between the variability ratio and the QPO centroid frequency measured
from these sources.

\section{Discussion and Conclusions}

\subsection{\lr\ of BHCs when QPO are Observed}

While there is no simultaneous spectral data for the six hour
observation period during which 0.08 Hz QPO were found in LMC~X-1
\cite{ebisawa89}, in the 13-hrs following this period (observed with
16-sec time resolution), the \lr\ was in the range 0.033-0.087
(Fig.~\ref{fig:colorccd}) which is very close to the \lr\ of
GS~1124-68 and GX~339-4 when QPO were observed. Although
High-Frequency PDS were produced by Ebisawa \etal, no 3-10 Hz QPO were
reported.  However, the QPO can turn on and off on a timescale of
\approx 1 sec \cite{miyamoto91,takizawa97}, correlated with a sudden
change in source spectral hardness (and intensity -- the so-called
``flip-flops''), and this may explain the absence of QPO in LMC~X-1.

We conclude that we observe no BHCs with spectral hardness
0.02~\approxlt~\lr~\approxlt~0.1 which do not exhibit QPO;
consequently, we expect that BHCs with spectral hardnesses in this
range should produce QPO similar to that described here. 

\subsection{The Very High State as a Transition between the Low State
and the High State}

Previous investigations using the present data concluded that both
band-limited noise variability behavior (like that seen in the Low
State) and power-law noise variability (like that seen in the High
State) are observed at various times simultaneous with QPO in GX~339-4
\cite{miyamoto91} and GS~1124-68 (T97).  In both studies, it
was shown that these behaviors are observed when the object was
relatively spectrally hard (in the case of Low State-like noise) and
spectrally soft (in the case of High State-like noise).  Neither
investigation commented upon the relationship between this transition
and the expected transition between the Low and High States.

In GS~1124-68 (T97), it was also found that the QPO centroid frequency
had (at least) two distinct functional dependencies on the 2-20 keV
intensity, one which was valid when power-law (HS-like PDS)
variability was observed (in Obs \#11 and \#15), the other when
band-limited (LS-like PDS) variability was observed. Comparison with
the present results shows that the change in the dependence of QPO
frequency on intensity observed by T97 occurs at \lr $\sim$0.055
(``the transition point'').

Because:
\begin{enumerate}
\item the Very High State of GX~339-4 and GS~1124-68 (\ie\ when QPO
are observed) is spectrally between the Low State and High State (0.02
\approxlt \lr\ \approxlt 0.2); 
\item power-law PDS, similar to that expected from the spectrally soft
High State, are observed when QPO are present in both GX~339-4 and
GS~1124-68 when these sources are spectrally softer than at the
transition point (0.02 \approxlt \lr \approxlt 0.055); in addition,
the QPO centroid frequency is correlated with spectral hardness in
this spectral region; 
\item band-limited PDS, similar to that expected from the spectrally
hard High State, are observed when QPO are present in both GX~339-4
and GS~1124-68 and when these sources are spectrally harder than at
the transition point (0.055 \approxlt \lr \approxlt 0.2); in addition,
the QPO centroid frequency is anti-correlated with spectral hardness
in this spectral region;
\end{enumerate}

\noindent it seems natural to interpret the VHS as the transition
between accretion dominated by the spherical corona (the Low-State)
and accretion dominated by the accretion disk (the High-State).  The
change at the (so-defined) transition point in the QPO spectral
hardness relative to the BBN variability, as well as the dependence of
the QPO centroid frequency on the \lr\ from a correlation to an {\it
anti-}correlation, may indicate that the QPO is a manifestation of the
interaction between the corona and the accretion disk.

It is widely assumed that the accretion rate (\mdot) is correlated
with the observed flux in BHCs, and that the 2-20 keV flux is used as
a monotonic indicator of this; thus, \mdot\ increases monotonically
from Off-State, to LS, to Intermediate State, to HS, to VHS
(cf. \citenp{esin97}).  Our results show that, when a BHC evolves
between LS and HS, it should cross the spectral region in which we
observe the VHS (being intermediate to the two), and that while doing
so, the noise properties of the BHC should change from band-limited to
power-law.  This has been observed recently: Cyg~X-1 exhibited 3-12 Hz
QPO during its 1996 transition, when its 2-20 keV intensity was
between that of observed HS and LS\cite{cui97}; and GRO J1655-40,
which exhibited 6.5 and 0.8 Hz QPO \cite{mendez98}. 

While Cyg~X-1 underwent its 1996 transition, QPO (of a broad type,
Q\approx 1) were observed, with centroid frequencies between 3-12 Hz
\cite{cui97}.  In Fig.~\ref{fig:cui}, we show the dependence of
Cyg~X-1 QPO $\nu_c$ vs. spectral hardness (from Table 2 and 3 of
\citenp{cui97}, and W. Cui, priv. comm.). The hardness ratio of
13.1-60 keV/2.0-6.5 keV was measured by {\it R}XTE/PCA, and cannot be
directly compared to the \lr\ measured by \ginga , although we can
reasonably expect that the two are monotonically related. 

Those observations with hardness ratio $>$0.11 had a type of
band-limited noise, with a power-law slope $\sim$0.3-0.8 below
$\sim$0.5 Hz, then constant between $\sim$0.5-2.0 Hz, breaking again
into a power-law of slope $\sim$ 2.0 above that; during this period
(called the ``transition phase'' by Cui \etal\ ) the QPO centroid
frequency was anti-correlated with spectral hardness.  During those
observations with hardness ratio $<$0.11 (called ``soft state'' by Cui
\etal\ ) the source variability was described as a broken power-law
(of slopes $\sim$ 1 and 2, breaking near $\nu \sim$ 5-15 Hz); in
addition, the QPO centroid frequency was below that observed in the
softest (but relatively spectrally harder) ``transition phase''
observations, indicating an anti-correlation with spectral hardness.

Comparison between the dependence on the QPO centroid frequency in
Cyg~X-1 (Fig.~\ref{fig:cui}) with that of GX~339-4 and GS~1124-68
(Fig.~\ref{fig:vhsqpo}a) shows that in all three objects, the centroid
frequency is correlated with spectral hardness, which changes to an
{\it anti-}correlation above some spectral hardness; in addition, all
three objects show power-law PDS behavior when the centroid frequency
is correlated with spectral hardness, and band-limited PDS behavior
when the centroid frequency is anti-correlated with spectral hardness.

The differences between the behavior of Cyg~X-1 and that which we
observe from GX~339-4 and GS~1124-68 is that the QPO (in Cyg X-1) are
considerably broader QPO (Q\approx 1 compared with Q\approx10), and the
2-20 keV flux is only moderately greater than its Low State flux, and
still below that of its High State \cite{belloni97}, while GX~339-4
and GS~1124-68 were observed with $\sim$ 2-20 keV intensities greater
than periods where ``High State'' behavior was observed. It was on the
basis of this discrepancy in 2-20 keV intensity in Cyg X-1
\cite{cui97}, as well as during a period when QPO were observed at low
intensity in GS~1124-68 \cite{belloni97}, that analogy with the Very
High State was discarded.  However, the change in dependence of the
QPO centroid frequency on spectral hardness, associated with the
change in the BBN variability is suggestive that the phenomena of
Cyg~X-1 observed during transition between Low and High state is the
same as that which we describe of GS~1124-68 and GX~339-4 in their
Very High State, in spite of the differences in $\sim$ 2-20 keV
intensity.  

We conclude that the conditions for transition between the two states
occur at luminosities both above (in the case of the Very High State)
and below (in the case of the Intermediate State) the High-State in
BHCs.

We cannot exclude that the assumption that the 2-20 keV flux (usually
thought to correspond to $\dot{M}$) is a monotonic indicator of states
is not correct.  For example, GS~2023+33 exhibited spectral and timing
behavior which has been classified as ``Low State'' (spectrally hard,
with band-limited noise), even at its highest luminosity -- although
much caution must be employed in interpreting this behavior, as
GS~2023+33 was a peculiar X-ray transient in more than one way
(cf. \citenp{tanakalewin95}).

The strength of the underlying BBN variability is (in GS~1124-68) on
average decreased by the presence of QPO.  One might interpret this to
mean that, while the physical process producing the VHS QPO is
distinct from that of the BBN variability, there is nonetheless
interaction between the two processes which affects the strength of
the BBN variability.   

It does not follow naturally from any other observed behavior that, as
a function of spectral hardness, the variability ratio should be low
($\sim$1), {\it increase} with increasing spectral hardness toward the
transition point (\lr\ $\sim$ 0.055, where it is \approxgt 4), and
then {\it decrease} with increasing spectral hardness. This therefore
represents a significant and interesting phenomenon, which requires
further discussion.  It implies that the spectrum of the QPO is (in
the Low State) similar to the spectrum of the BBN variability; as the
source spectrum evolves toward the High State, the QPO spectrum
becomes spectrally hard, relative to the BBN variability, becoming
maximally spectrally hard relative to the BBN variability at the same
\lr\ as when the QPO centroid frequency is at its maximum; and then, as the
spectrum evolves further toward the High State, the spectrum of the
QPO again softens toward the spectrum of the BBN variability. (We
emphasize here that ``evolution'' is as a function of spectral
hardness, not of time).  There exists no {\it a priori} reason that
the QPO photon spectrum should at any time be in any way similar to
the BBN variability photon spectrum -- the Variability Ratio could, in
principle, take on any value between 0 and infinity.  That the lowest
observed values of the Variability Ratio indicate that the QPO energy
spectrum is similar to the BBN energy spectrum suggests that the QPO
itself is due to the same mechanism as the BBN variability, which for
some unknown reason becomes enhanced at a particular frequency.

In summary, the observed dependence on spectral hardness of the BBN
variability while QPO are observed from GX~339-4 and GS~1124-68, and
their position on the color-color diagram spectrally intermediate to
the Low-State and High-State indicate that the so-called Very High
State QPO is due to interaction between the physical component
responsible for Low State behavior and that responsible for High State
behavior.  Our comparison with QPO observed from Cyg~X-1 at lower
luminosity, during transition between the Low and High States,
indicate that these QPO may also be due to such interaction.  If this
inferrence is correct, then the QPO may prove a useful probe in
investigating the properties of the intereaction of the different
accretion components in BHCs.

\label{sec:lsvhsqpo}

\label{sec:allconc}

\acknowledgements

RR is grateful to W. Cui for providing the unpublished spectral
hardness ratios measured from Cyg X-1.  RR is also grateful to
L. Bildsten for his hospitality at UC Berkeley where this work was
completed.  TO acknowledges an ESA Research Fellowship. MK gratefully
acknowledges the Visiting Miller Professor Program of the Miller
Institute for Basic Research in Science (UCB). WHGL acknowledges
support from NASA. 


\begin{thebibliography}{}

\bibitem[\protect\astroncite{{Abramowicz} {\rm et~al.\/}}{1995}]{abramowicz95}
{Abramowicz}, M.~A., {Chen}, X., \& {Taam}, R.~E., 1995,
\newblock {\em \apj} {\bf 452}, 379

\bibitem[\protect\astroncite{{Belloni} {\rm et~al.\/}}{1996}]{belloni96b}
{Belloni}, T., {M\'endez}, M., {Van Der Klis}, M., {Hasinger}, G., {Lewin}, W.
  H.~G., \& {Van Paradijs}, J., 1996,
\newblock {\em \apjl} {\bf 472}, L107

\bibitem[\protect\astroncite{{Belloni} {\rm et~al.\/}}{1997}]{belloni97}
{Belloni}, T., {Van Der Klis}, M., {Lewin}, W. H.~G., {Van Paradijs}, J.,
  {Dotani}, T., {Mitsuda}, K., \& {Miyamoto}, S., 1997,
\newblock {\em \aap} {\bf 322}, 857

\bibitem[\protect\astroncite{{Chen} {\rm et~al.\/}}{1995}]{chen95}
{Chen}, X., {Abramowicz}, M.~A., {Lasota}, J.-P., {Narayan}, R., \& {Yi}, I.,
  1995,
\newblock {\em \apjl} {\bf 443}, L61

\bibitem[\protect\astroncite{{Chen} \& {Taam}}{1996}]{chen96}
{Chen}, X. \& {Taam}, R.~E., 1996,
\newblock {\em \apj} {\bf 466}, 404

\bibitem[\protect\astroncite{{Cui}}{1997}]{cui97}
{Cui}, W., 1997,
\newblock {\em \apjl} {\bf 482}, L163

\bibitem[\protect\astroncite{Ebisawa}{1991}]{ebisawathesis}
Ebisawa, K., 1991,
\newblock {\em Ph.D. thesis}, University of Tokyo

\bibitem[\protect\astroncite{{Ebisawa} {\rm et~al.\/}}{1989}]{ebisawa89}
{Ebisawa}, K., {Mitsuda}, K., \& {Inoue}, H., 1989,
\newblock {\em \pasj} {\bf 41}, 519

\bibitem[\protect\astroncite{{Esin} {\rm et~al.\/}}{1997}]{esin97}
{Esin}, A.~A., {McClintock}, J.~E., \& {Narayan}, R., 1997,
\newblock {\em \apj} {\bf 489}, 865

\bibitem[\protect\astroncite{{Grebenev} {\rm et~al.\/}}{1991}]{grebenev91}
{Grebenev}, S.~A., {Siuniaev}, R.~A., {Pavlinskii}, M.~N., \& {Dekhanov},
  I.~A., 1991,
\newblock {\em Soviet Astronomy Letters} {\bf 17}, 985

\bibitem[\protect\astroncite{{Hasinger} \& {Van Der Klis}}{1989}]{hvdk89}
{Hasinger}, G. \& {Van Der Klis}, M., 1989,
\newblock {\em \aa} {\bf 225}, 79

\bibitem[\protect\astroncite{Lewin {\rm et~al.\/}}{1995}]{xrb}
Lewin, W., {Van Paradijs}, J., \& {Van Den Heuvel}, E. (eds.), 1995,
\newblock {\em X-Ray Binaries}, Vol.~1,
\newblock Cambridge University Press

\bibitem[\protect\astroncite{{Makishima} {\rm et~al.\/}}{1986}]{makishima86}
{Makishima}, K., {Maejima}, Y., {Mitsuda}, K., {Bradt}, H.~V., {Remillard},
  R.~A., {Tuohy}, I.~R., {Hoshi}, R., \& {Nakagawa}, M., 1986,
\newblock {\em \apj} {\bf 308}, 635

\bibitem[\protect\astroncite{{M\'endez} {\rm et~al.\/}}{1998}]{mendez98}
{M\'endez}, M., {Belloni}, T., \& {Van Der Klis}, M., 1998,
\newblock {\em \apjl},
\newblock in press

\bibitem[\protect\astroncite{{M\'endez} \& {Van Der Klis}}{1997}]{mendez97}
{M\'endez}, M. \& {Van Der Klis}, M., 1997,
\newblock {\em \apj} {\bf 479}, 926

\bibitem[\protect\astroncite{{Mitsuda} {\rm et~al.\/}}{1984}]{mitsudadiskbb}
{Mitsuda}, K., {Inoue}, H., {Koyama}, K., {Makishima}, K., {Matsuoka}, M.,
  {Ogawara}, Y., {Suzuki}, K., {Tanaka}, Y., {Shibazaki}, N., \& {Hirano}, T.,
  1984,
\newblock {\em \pasj} {\bf 36}, 741

\bibitem[\protect\astroncite{{Miyamoto} {\rm et~al.\/}}{1993}]{miyamoto93}
{Miyamoto}, S., {Iga}, S., {Kitamoto}, S., \& {Kamado}, Y., 1993,
\newblock {\em \apjl} {\bf 403}, L39

\bibitem[\protect\astroncite{{Miyamoto} {\rm et~al.\/}}{1991}]{miyamoto91}
{Miyamoto}, S., {Kimura}, K., {Kitamoto}, S., {Dotani}, T., \& {Ebisawa}, K.,
  1991,
\newblock {\em \apj} {\bf 383}, 784

\bibitem[\protect\astroncite{{Miyamoto} {\rm et~al.\/}}{1994}]{miyamoto94}
{Miyamoto}, S., {Kitamoto}, S., {Iga}, S., {Hayashida}, K., \& {Terada}, K.,
  1994,
\newblock {\em \apj} {\bf 435}, 398

\bibitem[\protect\astroncite{{Narayan}}{1996}]{narayan96}
{Narayan}, R., 1996,
\newblock {\em \apj} {\bf 462}, 136

\bibitem[\protect\astroncite{{Narayan} \& {Yi}}{1994}]{narayan94}
{Narayan}, R. \& {Yi}, I., 1994,
\newblock {\em \apjl} {\bf 428}, L13

\bibitem[\protect\astroncite{{Narayan} \& {Yi}}{1995}]{narayan95}
{Narayan}, R. \& {Yi}, I., 1995,
\newblock {\em \apj} {\bf 452}, 710

\bibitem[\protect\astroncite{Press {\rm et~al.\/}}{1995}]{press}
Press, W., Flannery, B., Teukolsky, S., \& Vetterling, W., 1995,
\newblock {\em Numerical Recipies in C},
\newblock Cambridge University Press

\bibitem[\protect\astroncite{Rutledge {\rm et~al.\/}}{1998}]{bhclfn98}
Rutledge, R., Lewin, W. H.~G., Dotani, T., Mitsuda, K., Paradijs, J.~V.,
  Vaughan, B., \& Klis, M. V.~D., 1998,
\newblock in progress

\bibitem[\protect\astroncite{Rutledge}{1996}]{rutledgethesis}
Rutledge, R.~E., 1996,
\newblock {\em Ph.D. thesis}, Massachusetts Institute of Technology

\bibitem[\protect\astroncite{{Takizawa} {\rm et~al.\/}}{1997}]{takizawa97}
{Takizawa}, M., {Dotani}, T., {Mitsuda}, K., {Matsuba}, E., {Ogawa}, M.,
  {Aoki}, T., {Asai}, K., {Ebisawa}, K., {Makishima}, K., {Miyamoto}, S.,
  {Iga}, S., {Vaughan}, B., {Rutledge}, R.~E., \& {Lewin}, W. H.~G., 1997,
\newblock {\em \apj} {\bf 489}, 272

\bibitem[\protect\astroncite{Tanaka \& Lewin}{1995}]{tanakalewin95}
Tanaka, Y. \& Lewin, W., 1995,
\newblock in \cite{xrb}, p. 126

\bibitem[\protect\astroncite{Turner {\rm et~al.\/}}{1989}]{turner89}
Turner, M., Thomas, H., Patchett, B., Reading, D., Makishima, K., Ohashi, T.,
  Dotani, T., Hayashida, K., Inoue, H., Kondo, H., Koyama, K., Mitsuda, K.,
  Ogawara, Y., Takano, S., Awak, H., Tawara, Y., \& Nakamura, N., 1989,
\newblock {\em \pasj} {\bf 41}, 345

\bibitem[\protect\astroncite{{Van Der Klis}}{1994a}]{mvdk94a}
{Van Der Klis}, M., 1994a,
\newblock {\em \aa} {\bf 283}, 469

\bibitem[\protect\astroncite{{Van Der Klis}}{1994b}]{mvdk94b}
{Van Der Klis}, M., 1994b,
\newblock {\em \apjs} {\bf 92}, 511

\bibitem[\protect\astroncite{{Van Der Klis}}{1995}]{mvdk95}
{Van Der Klis}, M., 1995,
\newblock in \cite{xrb}, p. 252

\bibitem[\protect\astroncite{{Van Paradijs}}{1995}]{jan95}
{Van Paradijs}, J., 1995,
\newblock in \cite{xrb}, p. 536

\end{thebibliography}

\newpage

\begin{table}[htb]
\begin{center}
\caption{Quantity of BHC Data for Timing Analysis, by Source and
Temporal Resolution}
\label{tab:timingdata}
\vspace{0.1cm}
\begin{tabular}{lcrrrrr} \hline \hline
  	     &  Time Res. (sec)   &\multicolumn{1}{c}{0.00781} &\multicolumn{1}{c}{  0.06250}    &\multicolumn{1}{c}{ 0.5000}     &\multicolumn{1}{c}{$>$0.5}  &\multicolumn{1}{c}{Total}   \\ 
Source       &     &\multicolumn{1}{c}{(ksec)} &\multicolumn{1}{c}{       (ksec)} &\multicolumn{1}{c}{(ksec)}   &\multicolumn{1}{c}{(ksec)} &\multicolumn{1}{c}{(ksec)}   \\ \hline
Cyg X-1      &     &   25.2    &     4.2      &   1.9      & 55.8  &    87.0 \\
GS 1124-68   &     &   54.6    &    57.9      &  268.4     &  0.0  &   380.8 \\
GS 2000+25   &     &   30.7    &     9.4      &   27.1     & 71.2  &   138.3 \\
GX 339-4     &     &   15.2    &    23.0      &   15.2     & 18.9  &    72.4 \\
\hline                                                          
\end{tabular}
\end{center}
\end{table}

\begin{table}[htb]
\begin{center}
\begin{normalsize}
\caption{QPO Parameters for Best-Fit Cyg X-1 Composite PDS}
\label{tab:cygx1compositepdsqpoparams}
\vspace{0.1cm}
\begin{tabular}{ccccc} \hline \hline
Spectral Range & Energy Range  & $\nu_c$         &  FWHM              &  \%rms           \\
 (\lr)         & (keV)         & (Hz)          &  (Hz)              &                  \\ \hline
G1$^a$         &2.3--4.6       & 1.07\ppm0.04  & 1.30\ppm 0.09      & 11.7\ppm 0.5     \\
(0.1--0.15)    &4.6--9.2       & 1.15 \ppm 0.03& 1.04 \ppm0.08      & 10.1\ppm 0.4     \\
               &9.2--18.4      & 1.14\ppm 0.03 & 0.8  \ppm 0.1      & 8.2  \ppm 0.6    \\
               &2.3--18.4      & 1.11 \ppm 0.03& 1.17 \ppm 0.07     & 10.8 \ppm 0.4    \\  \hline
G2$^b$         &2.3--4.6       & (3.14)        & (0.72)             & 3.5\ud{0.3}{0.4} \\  
 (0.15--0.32)  &4.6--9.2       & (3.14)        & (0.72)             & 3.3\ud{0.5}{0.6} \\
               &9.2--18.4      & (3.14)        & (0.72)             & $<$6.0           \\
               &2.3--18.4      & 3.14\ud{0.09}{0.07}&0.7\ppm 0.3    & 3.5\ud{0.3}{0.4} \\  \hline
\freqrange{5}{0.001}{64}{$^a$} \\
\freqrange{5}{0.0156}{64}{$^b$} \\
\fixedval{5} \\
\upperlim{5}\\
\end{tabular}
\end{normalsize}
\end{center}
\end{table}


\begin{table}[htb]
\begin{center}
\begin{scriptsize}
\caption{\%rms of QPO and BBN in GX~339-4}
\label{tab:gx339vhsrms}
\vspace{0.1cm}
\begin{tabular}{cccccccc} \hline \hline
               & Energy Range &  QPO  \%rms   & BBN \%rms         & BBN  	  &    FWHM     &   $\nu_c $            &  $\chi^2_\nu$ (dof)   \\
\lr            &  (keV)       &               &   (0.001-64 Hz)  & model  &    (Hz)     &    (Hz)               &                       \\ \hline               
0.030--0.043   & 2.3--4.6     & 1.79\ppm 0.04 & 3.89\ppm 0.09    & PL     &0.53\ppm 0.03& 5.80\ppm 0.02         & 1.36 (139)             \\
(F1)           & 4.6--9.2     & 5.97\ppm 0.08 & 6.3\ppm 0.2      & PL      &0.53\ppm 0.02& 5.80\ppm 0.01         & 1.87 (139)            \\
               & 9.2--18.4    & 10.5\ppm 0.2  & 5.6\ppm 0.5      & PL      &0.55\ppm 0.03& 5.78\ppm 0.01         & 1.36 (140)            \\
               & 2.3--18.4    & 2.89\ppm 0.04 & 4.45\ppm 0.07    & PL      &0.47\ppm 0.02& 5.80\ppm 0.01         & 2.42 (139)            \\  \hline
0.043--0.050   & 2.3--4.6     & 2.29\ud{0.3}{0.08} & 13.5\ppm 0.8& DFTPL   &0.59\ud{0.2}{0.3}& 6.40\ppm 0.04     & 2.25 (46)             \\       
(F2)           & 4.6--9.2     & 8.7\ppm 0.1   & 23.3\ppm 1.5     & DFTPL   &1.07\ppm 0.06& 6.37\ppm 0.02         & 2.60 (46)            \\           
               & 9.2--18.4    & 17.4\ppm 0.4  & 24.3 \ppm 2      & DFTPL   & 1.4\ppm 0.1 & 6.42\ppm 0.04         & 1.58 (49)            \\        
               & 2.3--18.4    & 4.24\ppm 0.06 & 15.4\ppm 0.7     & DFTPL   &0.95\ppm 0.05& 6.36\ppm 0.02         & 4.28 (46)            \\  \hline
0.050--0.060   & 2.3--4.6     & 1.1\ppm 0.2   & 8.4\ppm 0.2      & FTPL    & (3.58)      & (6.9)                 & 1.13 (48)             \\       
(F3)           & 4.6--9.2     & 5.7\ppm 0.1   & 13.0\ppm 0.2     & FTPL     & 3.17\ppm 0.2& 7.08\ppm 0.07         & 1.43 (47)            \\        
               & 9.2--18.4    & 11.5\ppm 0.4  & 8.9\ppm 0.7      & FTPL     & 3.55\ppm 0.3& 7.38\ppm 0.09         & 1.05 (48)            \\        
               & 2.3--18.4    & 2.7 \ppm 0.1  & 9.5\ppm 0.1      & FTPL     &3.40\ppm 0.3 & 6.9 \ppm 0.1          & 1.39 (46)            \\  \hline 
0.060--0.075   & 2.3--4.6     & 1.85\ppm 0.06 & 8.8\ppm 0.1      & FTPL     &(2.27)       & (6.27)                & 2.73 (47)             \\        
(F4)           & 4.6--9.2     & 5.9\ppm 0.1   & 13.2\ppm 0.3     & FTPL     & 2.35\ppm 0.1& 6.38\ppm 0.03         & 2.12 (45)            \\         
               & 9.2--18.4    & 9.3\ppm 0.2   & 8.0\ppm 0.4      & FTPL     & 1.80 \ppm 0.1 & 6.39\ppm 0.04       & 2.30 (47)            \\         
               & 2.3--18.4    & 3.24\ppm 0.05 & 9.6\ppm 0.1      & FTPL    &2.37\ppm 0.09& 6.27\ppm  0.03        & 2.71 (45)            \\  \hline 
\freqrange{5}{0.125}{64}{$$} \\
\end{tabular}
\end{scriptsize}
\end{center}
\end{table}

\begin{table}[htb]
\begin{small}
\begin{center}
\caption{Division of GS~1124-68 Data }
\label{tab:g1124takidata}
\vspace{0.1cm}
\begin{tabular}{lrrcc} \hline \hline
Observation  &\multicolumn{1}{c}{Start Time }  & \multicolumn{1}{c}{  End Time}   & PDS              &               \\
ID           &\multicolumn{1}{c}{  (UT)    }     & \multicolumn{1}{c}{  (UT)  }     & Freq Range       & $<$\lr$>$   \\ \hline 
1            & 91/011 00:51  &  01:17       &  0.0156--8.0 Hz  & 0.155 \ppm  0.004  \\
2            & 91/011 19:37  &  20:42       &  0.0156--8.0     & 0.097 \ppm   0.006  \\
3            & 91/011 20:49  &  91/012 00:05&  0.0156--64.0    & 0.092 \ppm   0.003  \\
4 (not used) &\multicolumn{1}{c}{  --    }       & \multicolumn{1}{c}{ -- }         &   --             & -- \\
5            & 91/014 19:48  &  23:25       &  0.0156--8.0     & 0.054  \ppm  0.004   \\
6            & 91/016 18:46  &  23:27       &  0.0156--8.0     & 0.026 \ppm   0.005   \\
7            & 91/017 18:05  &  21:57       &  0.0156--8.0     & 0.014 \ppm   0.001   \\
8            & 91/018 17:30  &  21:27       &  0.0156--8.0     & 0.024 \ppm   0.003   \\
9            & 91/020 15:17  & 17:15        &  0.0156--8.0     & 0.027 \ppm   0.001   \\
10           & 91/022 17:18  & 20:09        &  0.0156--64.0     & 0.058 \ppm   0.004   \\
11a          & 91/025 12:48  & 13:10        &  0.0156--64.0     & 0.036 \ppm   0.001   \\
11b          & 91/025 13:46  & 13:56        &  0.0156--64.0     & 0.044 \ppm   0.001   \\
12           & 91/036 10:34  & 11:07        &  0.0156--64.0     & 0.027 \ppm   0.002   \\
13           & 91/037 10:05  & 12:13        &  0.0156--64.0     & 0.029 \ppm   0.001   \\
14           & 91/044 05:04  & 08:38        &  0.0156--8.0     & 0.024 \ppm   0.004   \\
15           & 91/045 06:11  & 08:10        &  0.0156--64.0     & 0.013 \ppm   0.001   \\ 
16           & 91/051 23:29  & 91/053 04:32 &  0.0156--64.0    & 0.0093 \ppm 0.0017   \\ 
17           & 91/056 20:56  & 91/058 00:19 &  0.0156--64.0    & 0.0068 \ppm 0.0007   \\ 
18           & 91/067 14:28  & 91/109 20:13 &  0.0156--64.0    & 0.0009 \ppm 0.0007   \\ 
19           & 91/137 03:12  & 91/204 21:04 &  0.0156--64.0    & 0.3995 \ppm 0.1578   \\ \hline
\end{tabular}
\end{center}
\end{small}
\end{table}

\begin{table}[htb]
\begin{center}
\begin{scriptsize}
\caption{GS 1124-68 QPO Fundamental Parameters}
\label{tab:g1124takiparams}
\vspace{0.1cm}
\begin{tabular}{lccccc} \hline \hline
Observation  & Energy Range  &  $\nu_c$       &  FWHM         &  \% rms            & \chisqrnu\ (dof) \\
ID           &  (keV)        &   (Hz)         &  (Hz)         &                    &                 \\ \hline
1            &  2.3--4.6     & 3.004\ppm 0.008& 0.37\ppm 0.1  & 2.4 \ppm 0.1       & 1.61 (39) \\
             &  4.6--9.2     & 2.98 \ppm0.01  & 0.36\ppm 0.05 & 4.9 \ppm 0.2       & 1.24 (39) \\
             &  9.2--18.4    & 3.004\ppm0.008 & 0.29\ppm 0.05 & 7.0 \ppm 0.1       & 2.46 (39) \\
             &  2.3--18.4    & 2.98\ppm0.001  & 0.50\ppm 0.06 & 4.0 \ppm 0.1       & 1.24 (39)   \\ \hline
2            &  2.3--4.6     & 5.02\ppm 0.05  & 1.02\ppm 0.2  & 1.6 \ppm 0.2       & 1.20 (41)  \\
             &  4.6--9.2     & 5.07\ppm 0.02  & 0.80\ppm 0.04 & 4.8\ppm 0.1        & 1.99 (43)       \\      
             &  9.2--18.4    & 5.02\ppm 0.02  & 0.69\ppm 0.04 & 7.9\ppm 0.2        & 1.99 (43)  \\
             &  2.3--18.4    & 5.03\ppm 0.02  & 0.83\ppm 0.05 & 2.86\ppm 0.08      & 1.80 (41)  \\ \hline
3            &  2.3--4.6     & 5.33\ppm 0.02  & 0.97\ppm 0.06 & 1.74\ppm 0.05      & 1.33 (500) \\
             &  4.6--9.2     & 5.325\ppm 0.008& 0.69\ppm 0.03 & 5.4 \ppm 0.1       & 1.34 (500) \\
             &  9.2--18.4    & 5.33\ppm 0.009 & 0.74\ppm 0.03 & 10.6\ppm 0.1       & 1.32 (509) \\
             &  2.3--18.4    & 5.329\ppm 0.007& 0.82\ppm 0.03 & 3.19\ppm 0.05      & 1.80 (501) \\ \hline
5            &  2.3--4.6     & (7.38)         & (2.46)        & $<$0.80            & 1.56 (43) \\
             &  4.6--9.2     & 6.82\ppm 0.09  & 1.7\ud{0.4}{0.6}& 2.4\ppm 0.4      & 1.64 (41) \\
             &  9.2--18.4    & (7.38)         & (2.46)        & 9.2\ppm 0.2        & 1.97 (46)\\
             &  2.3--18.4    & 7.4\ppm 0.1    & 2.45\ppm 0.3  & 1.3 \ppm 0.14      & 2.29 (41) \\ \hline
8	     &  2.3--4.6     & (5.65)         &  (2.2)        & 0.9\ppm 0.07       & 1.65 (47) \\
             &  4.6--9.2     & (5.65)         &  (2.2)        & 2.7\ppm 0.1        & 1.95 (47) \\
             &  9.2--18.4    & (5.65)         &  (2.2)        & 8.8\ppm 0.6        & 2.00 (47) \\
             &  2.3--18.4    & 5.65\ppm 0.15  & 2.2\ud{0.9}{0.6}& 1.0\ppm 0.04     & 0.99 (42)  \\ \hline
10           &  2.3--4.6     & (7.72)         &  (3.7)        & 0.7\ppm 0.1        & 1.64 (46) \\
             &  4.6--9.2     & 8.11 \ppm 0.05 & 3.9  \ppm 0.2 & 6.0 \ppm 0.3       & 1.68 (44) \\
             &  9.2--18.4    & 8.43 \ppm 0.03 & 3.27\ppm 0.08 & 16.4\ppm 0.2       & 1.73 (45)  \\
             &  2.3--18.4    & 7.72 \ppm 0.04 & 3.7 \ppm 0.1  & 2.55 \ppm 0.06     & 2.27 (44) \\  \hline
11a          &  2.3--4.6     & (7.5)          & (2.11)         & $<$0.36           & 1.27 (52) \\
             &  4.6--9.2     & (7.5)          & (2.11)         & 1.4\ppm 0.2       & 1.75 (50) \\
             &  9.2--18.4    & (7.5)          & (2.11)         & $<$5.1            & 0.76 (52) \\ 
             &  2.3--18.4    & 7.8\ppm 0.3    & 2.11\ppm 0.6  & 0.56\ppm 0.07      & 2.55 (52)  \\\hline 
11b          &  2.3--4.6     & (5.82)         & (0.35)        & 1.10\ppm 0.08      & 1.47 (48) \\  
             &  4.6--9.2     & 5.816\ppm 0.009& 0.33\ppm 0.08 & 8.0\ud{0.8}{0.4}   & 1.06 (46) \\
             &  9.2--18.4    & 5.81\ppm 0.01  & 0.36 \ppm 0.15& 15.7\ud{3}{1}      & 0.72 (47)  \\ 
             &  2.3--18.4    & 5.813\ppm 0.008& 0.36\ppm 0.07 & 3.2\ppm 0.3        & 1.36 (46) \\ \hline
15           &  2.3--4.6     & 4.5 \ppm 0.1   & 1.3\ud{0.3}{0.2}& 0.67\ppm 0.05    & 1.20 (46) \\ 
             &  4.6--9.2     & 4.83\ppm 0.09  & 1.1\ud{0.3}{0.2}& 1.6\ppm 0.2      & 1.22 (39) \\
             &  9.2--18.4    & (4.68)         & (1.14)        & $<$4.8             & 1.58 (51) \\
             &  2.3--18.4    & 4.68\ppm 0.04  & 0.96 \ppm 0.14  & 0.75\ppm 0.04      & 1.36 (39) \\ \hline
\multicolumn{6}{l}{Frequency ranges used to fit and \lr\ are in Table~\protect{\ref{tab:g1124takidata}}} \\
\fixedval{6} \\
\upperlim{6} \\
\end{tabular}
\end{scriptsize}
\end{center}
\end{table}


\begin{table}[htb]
\begin{center}
\begin{scriptsize}
\caption{QPO Parameters for GS 2000+25}
\label{tab:gs2000qpopdsparams}
\vspace{0.1cm}
\begin{tabular}{cccccccc} \hline \hline
            &  \multicolumn{3}{c}{QPO}                 & &                 \\ \cline{2-4} \cline{6-8}
Energy Range&  $\nu_c$    & FWHM        & \%rms        & & \chisqrnu\ (dof) \\ 
 (keV)      &  (Hz)       & (Hz)        &              & &                 \\  \hline
2.3--4.6    & --          & --          & --           & & 3.40 (29)       \\
            & 2.41\ud{0.15}{0.09}&1.3\ud{0.6}{0.3}&4.9\ud{1.3}{0.7}& & 0.79(26) \\
4.6--9.2    & (2.63)      & (1.68)      & 5.6 \ppm 0.7 & & 1.52 (28)       \\
9.2--18.4   & (2.63)      & (1.68)      & 5.4\ud{1.3}{1.7}& & 1.91 (30)   \\
2.3--18.4   & 2.63\ppm 0.06& 1.7\ppm 0.2 & 6.3\ppm 0.5  & & 1.39 (26)      \\  \hline
\multicolumn{8}{l}{Fit is of single observation period with \lr=0.09-0.1} \\
\fixedval{8} \\
\freqrange{8}{0.125}{8.0}{$$}
\end{tabular}
\end{scriptsize}
\end{center}
\end{table}

\clearpage


\begin{figure}[htb]
\caption{\label{fig:colorccd}
  {\bf Color-Color Diagram}. 
	A composite color-color diagram of  all 10 BHC sources
  observed by \ginga\ (\tensources ), which
  shows the range of spectral hardness observed from each source.
  Data has been corrected for background counts as well as for aspect
  and deadtime, prior to producing the hardness ratios. Each
  point represents 128sec integration.  Only values of the hardness
  ratio which are $>$4\sig\ significant are shown.  
}
\end{figure}


\begin{figure}[htb]
\caption{\label{fig:qpopresentbig} {\bf Very High State and Low State
  QPO, Located on the \lr\ vs. \hr\ CCD}.  The position on the \lr\
  vs. \hr\ CCD, when QPO is observed from GS~1124-68 (Very High
  State), GX~339-4 (Very High State and Low State from Grebenev \etal\
  1991), Cyg X-1 (Low State), and GS~2000+25 (Low State).  For
  clarity, only a few representative points for each source are shown.
  This figure may be directly compared with the full composite CCD in
  Fig.~\protect{\ref{fig:colorccd}}.  While the black hole candidates
  occupy four orders of magnitude in \lr\ parameter space, the QPO are
  observed largely within one order of magnitude in \lr . }
\end{figure}


\begin{figure}[htb]
\caption{\label{fig:hrsources}
  {\bf Source Hardness Ratios}. 
	The \lr\ of all 10 BHC sources in the present study, which
  shows the range of spectral hardness observed from each source.  Each
  point represents 128 sec integration.  
}
\end{figure}


\begin{figure}[htb]
\caption{\label{fig:speclr}
  {\bf Intrinsic Source Energy Spectrum vs. \lr}. The intrinsic
  source photon spectrum is indicated by a combination of the ratio of
  the power-law photon flux to the black body photon flux (2-20 keV),
  the photon power-law slope ($\alpha$=0.9 to 2.9, spaced by 0.4), and
  the black-body temperature (0.1, 0.25, 0.4, 0.55, 0.7, 0.85, 1.0,
  and 1.4 keV).  The \lr\ is obtained by folding the intrinsic source
  photon spectrum through the \ginga\ LAC response matrix.  
}
\end{figure}


\begin{figure}[htb]
\caption{\label{fig:cygx1splitg}
  {\bf Cyg X-1 Composite PDS, G1 and G2}.  8 PDS of Cyg X-1, in 4
  energy ranges (indicated in each panel).  There are two PDS for each
  energy range: one from data taken when the source \lr\ is in the
  range 0.1--0.15 (G1; connected by a
  line to guide the eye), and one from data taken when the source \lr\ 
  is in the range 0.15--0.32 (G2, unconnected points). The 1\sig\ error bars are included.  The
  spectrally soft PDS shows a QPO feature near 1 Hz, while the harder
  PDS does not have this feature.  The PDS show roughly equal \%rms
  values above 10 Hz, and \%rms values which are not equal below 10
  Hz.  }
\end{figure}

\clearpage
\begin{figure}[htb]
\caption{ {\bf Detailed HID and PDS of GX 339-4 with QPO} Each column
  is headed with the energy range of the data used for those PDS, and
  each row is labeled according to spectrum, corresponding to the
  divisions in the HID at the top of the figure.  {\it Upper left-hand
  panel} is the \lr\ vs. time, during the period in which QPO were
  observed from GX~339-4 (Sept 1988; time=0 is 88/247 13:37 UT).  {\it
  Upper right-hand panel} is the HID, \ir\ \ccr\ vs. \lr\ during the
  Sept 1988 observation.  The panel has been divided by \lr\ value,
  and each region is labeled. {\it Bottom Panels}: The PDS in three
  energy ranges plus the intensity range (labeled by column) of the
  Sept 1988 observations. Upper-limits are 2\sig.  The data was
  divided by spectral hardness (each row is labeled, and corresponds
  to a label in the HID above).  The BBN PDS characteristics evolve --
  as the spectrum softens, the BB PDS changes from a
  flat-top-power-law to a pure power-law, while the QPO remains
  throughout.).  }
\label{fig:gx339qpohidpds}
\end{figure}

\clearpage
\begin{figure}[htb]
\caption{
  {\bf Detailed HID and PDS of GS~1124-68 with QPO} {\it Upper left-hand
    panel} is the \lr\ vs. time, during the first 40 days of outburst.  {\it
    Upper right-hand panel} is the HID, \ir\ \ccr\ vs. \lr .  The
  panel has been divided (identically to
  Fig.~\protect{\ref{fig:gx339qpohidpds}} by \lr\ value,
  and each region is labeled. {\it Bottom Panels}: The PDS in three
  energy ranges plus the intensity range (labeled by column) of the
  Sept 1988 observations. Upper-limits are 2\sig.  The data was divided by spectral hardness
  (each row is labeled, and corresponds to a label in the HID above).
  The BBN PDS characteristics evolve -- as the spectrum softens, the BB
  PDS changes from a flat-top-power-law to a pure power-law, while the
  QPO remains throughout.).
}
\label{fig:gs1124qpohidpds}
\end{figure}


\begin{figure}[htb]
\caption{
  {\bf GS 2000+25 PDS of State with QPO}. 
  The  PDS in the three energy ranges and the sum energy
  range (indicated in each panel).  Corrections for background and
  deadtime have been applied to the Power.   Upper-limits are 2\sig. 
  }
\label{fig:g2000qpopds}
\end{figure}


\begin{figure}[htb]
\caption{ {\bf QPO Centroid Frequency ($\nu_c$) and FWHM vs. \lr }.
  {\it Panel a}: Centroid frequency of the fundamental QPO, as a
  function of \lr .  Centroid frequency is maximum near \lr=0.055.
  {\it Panel b}. The FWHM of the fundamental vs. \lr.  The FWHM is
  greatest near \lr=0.055.  1\sig error bars are included on all
  points.  The Cyg~X-1 G2-data are not depicted (due to their high
  \lr) but follow these trends.  }
\label{fig:vhsqpo}
\end{figure}


\begin{figure}[htb]
\caption{
  {\bf QPO and BBN \%rms, Very High State QPO of GS~1124-68 \& GX~339-4}. 
  The \%rms variability in the QPO and BBN components when QPO is
  present or absent, in GS~1124-68 and GX~339-4.  Upper-limits are
  3-sigma.  Points for GS~1124-68 are marked with observation IDs from
  Table~\protect{\ref{tab:g1124takidata}}; points from GX~339-4 are marked with
  spectral indicator IDs taken from
  Fig.~\protect{\ref{fig:gx339qpohidpds}}  {\it (a)}: QPO  \%rms   vs. Energy range from GS~1124-68. 
  {\it (b)}: BBN  \%rms (when QPO is present)  vs. Energy range from
  GS~1124-68. 
  {\it (c)}: BBN  \%rms (when QPO is absent)  vs. Energy range from
  GS~1124-68. 
  {\it (d)}: QPO  \%rms   vs. Energy range from GX~339-4. 
  {\it (e)}: BBN  \%rms (when QPO is present)  vs. Energy range from
  GX~339-4.
  }
\label{fig:vhsbbqporms}
\end{figure}


\begin{figure}[htb]
\caption{
  {\bf \%rms of Low-State QPO vs. Energy Range, Cyg~X-1 and
    GS~2000+25}.  Cyg X-1 (open circles) and GS~2000+25 (solid
  squares) Error bars on the \%rms are 1\sig, while on energy they
  represent the energy range of the PDS.  The QPO \%rms
  is soft (in Cyg X-1), and flat (in GS~2000+25).  This contrasts with
  QPO in GS~1124-68 which, in the same spectral range (Obs. \#1-3),
  as well as in all other obseravations and in GX~339-4 as well, is  spectrally hard.  }
\label{fig:lsqpo}
\end{figure}


\begin{figure}[htb]
\caption{\label{fig:rrvswr}
  {\bf Variability Ratio vs. Spectral Hardness }. 
The Variability Ratio of QPO \%rms (Eq.~\ref{eq:rr}) as a
  function of \lr , from the four QPO sources in the present study.  
   A Variability
  Ratio of $X$ indicates the QPO is $X$ times  as strong relative to the
  BBN variability in the \her\ as it is in the \ler .  }
\end{figure}


\begin{figure}[htb]
\caption{\label{fig:rrvsvc}
  {\bf Variability Ratio vs. QPO Centroid Frequency}. 
The normalized ratio of QPO \%rms (Eq.~\ref{eq:rr}) vs. the QPO
  centroid frequency of sources in the present study.  A Variability
  Ratio of $X$ indicates the QPO is $X$ times  as strong relative to the
  BBN variability in the \her\ as it is in the \ler .  
  }
\end{figure}


\begin{figure}[htb]
\caption{\label{fig:cui} {\bf Cyg X-1 1996 Transition, QPO Centroid
  Frequency vs. Spectral Hardness}.  The $\nu_c$ are taken from
  Table 2 and 3 of Cui \etal\ (1997), and spectral hardnesses by
  W. Cui (priv. comm.).  The value of the spectral hardness ratio is
  as measured by {\it R}XTE/PCA and cannot be directly compared to the
  \lr\ measured by \ginga\ although the spectral hardness is
  monotonic, and so can be qualitatively compared to the $\nu_c$
  vs. \lr\ of GX~339-4 and GS~1124-68
  (Fig.~\protect{\ref{fig:vhsqpo}}).  Each point
  is marked with the observation number taken from the Table of Cui
  \etal\ (1997).  }
\end{figure}


\clearpage
\pagestyle{empty}
\begin{figure}[htb]
\PSbox{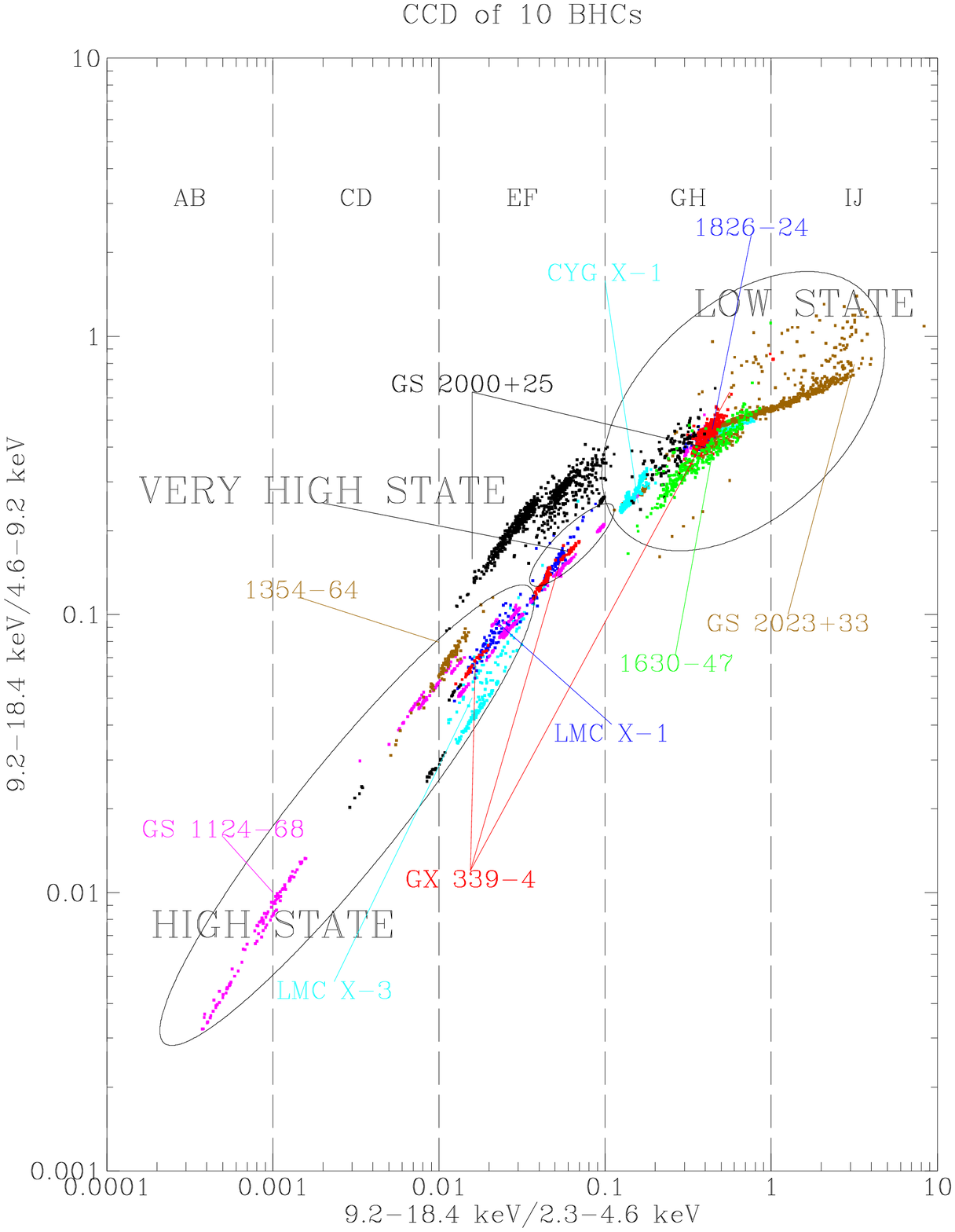 hoffset=-80 voffset=-80}{14.7cm}{21.5cm}
\FigNum{\ref{fig:colorccd}}
\end{figure}


\clearpage
\pagestyle{empty}
\begin{figure}[htb]
\PSbox{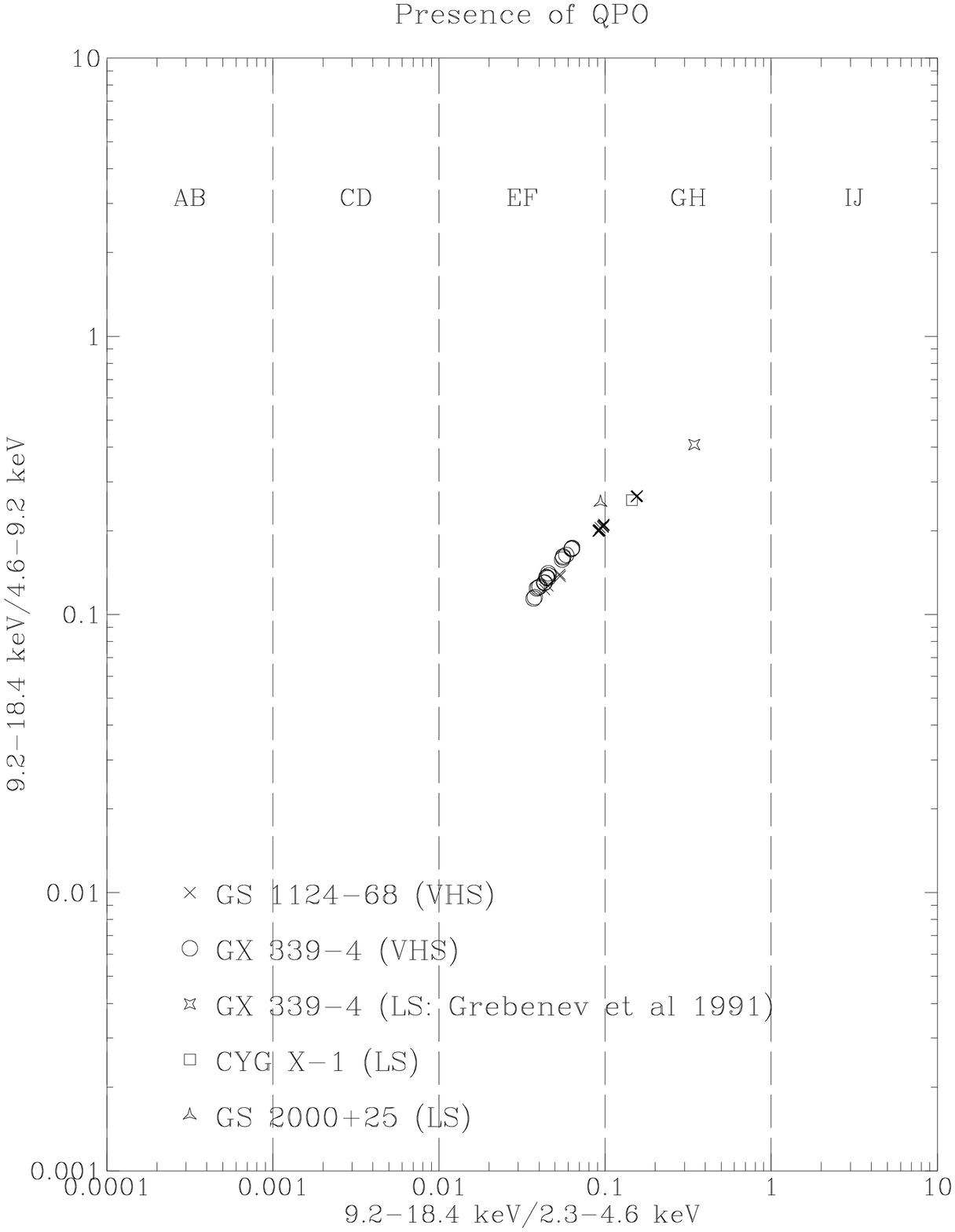 hoffset=-80 voffset=-80}{14.7cm}{21.5cm}
\FigNum{\ref{fig:qpopresentbig}}
\end{figure}


\clearpage
\pagestyle{empty}
\begin{figure}[htb]
\PSbox{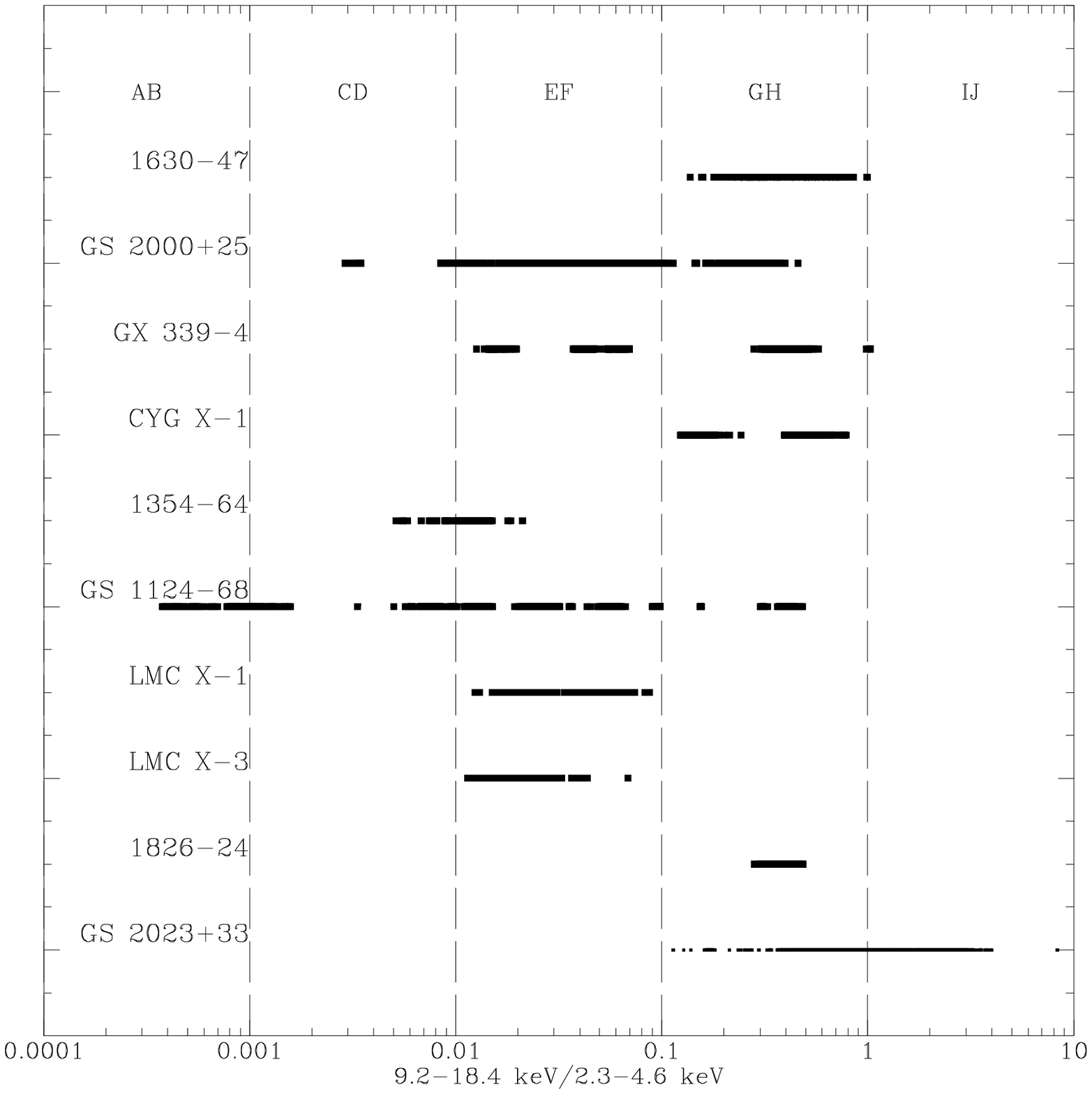  hoffset=-80 voffset=-80}{14.7cm}{21.5cm}
\FigNum{\ref{fig:hrsources}}
\end{figure}


\clearpage
\pagestyle{empty}
\begin{figure}[htb]
\PSbox{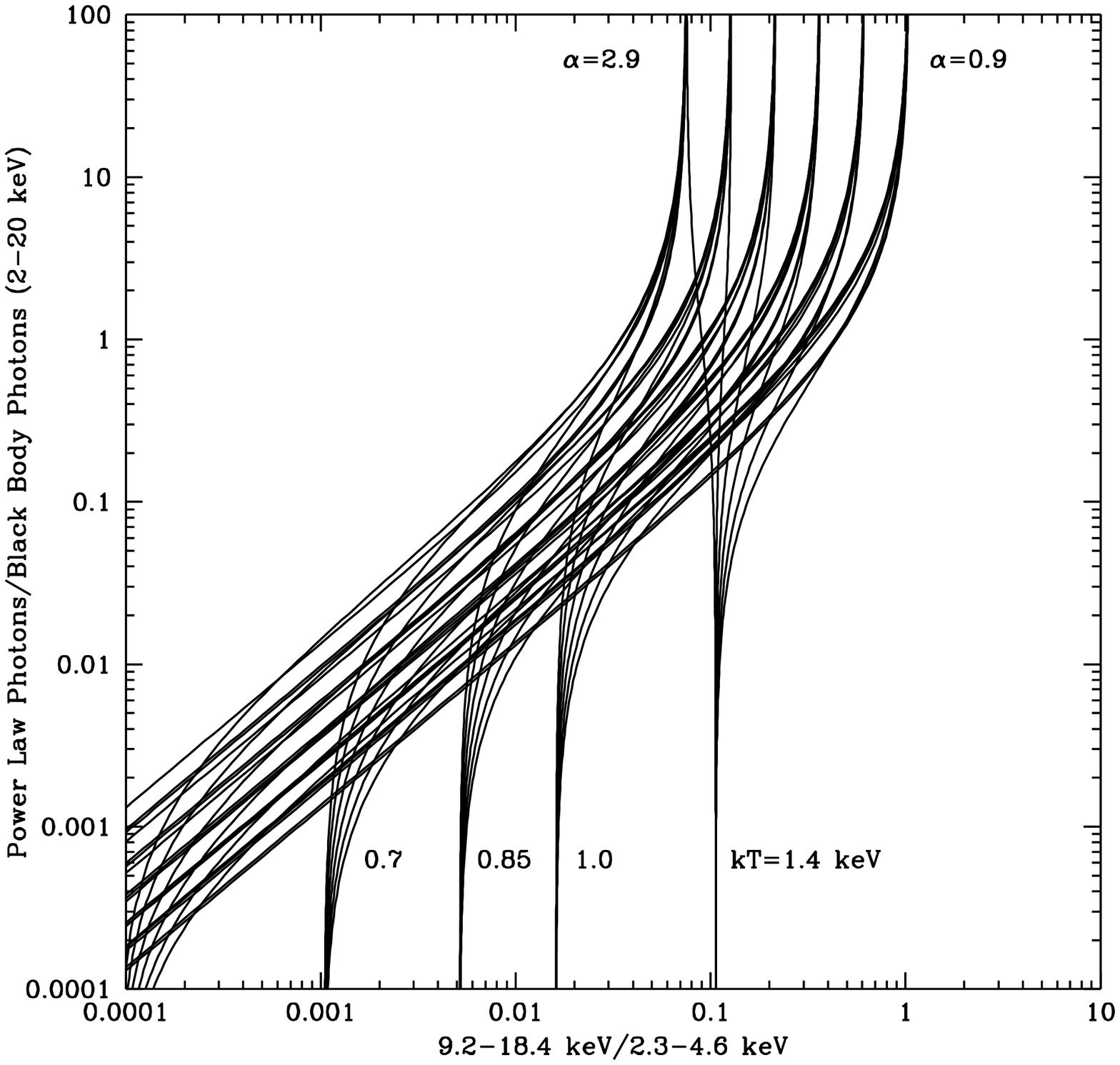  hoffset=-80 voffset=-80}{14.7cm}{21.5cm}
\FigNum{\ref{fig:speclr}}
\end{figure}


\clearpage
\pagestyle{empty}
\begin{figure}[htb]
\PSbox{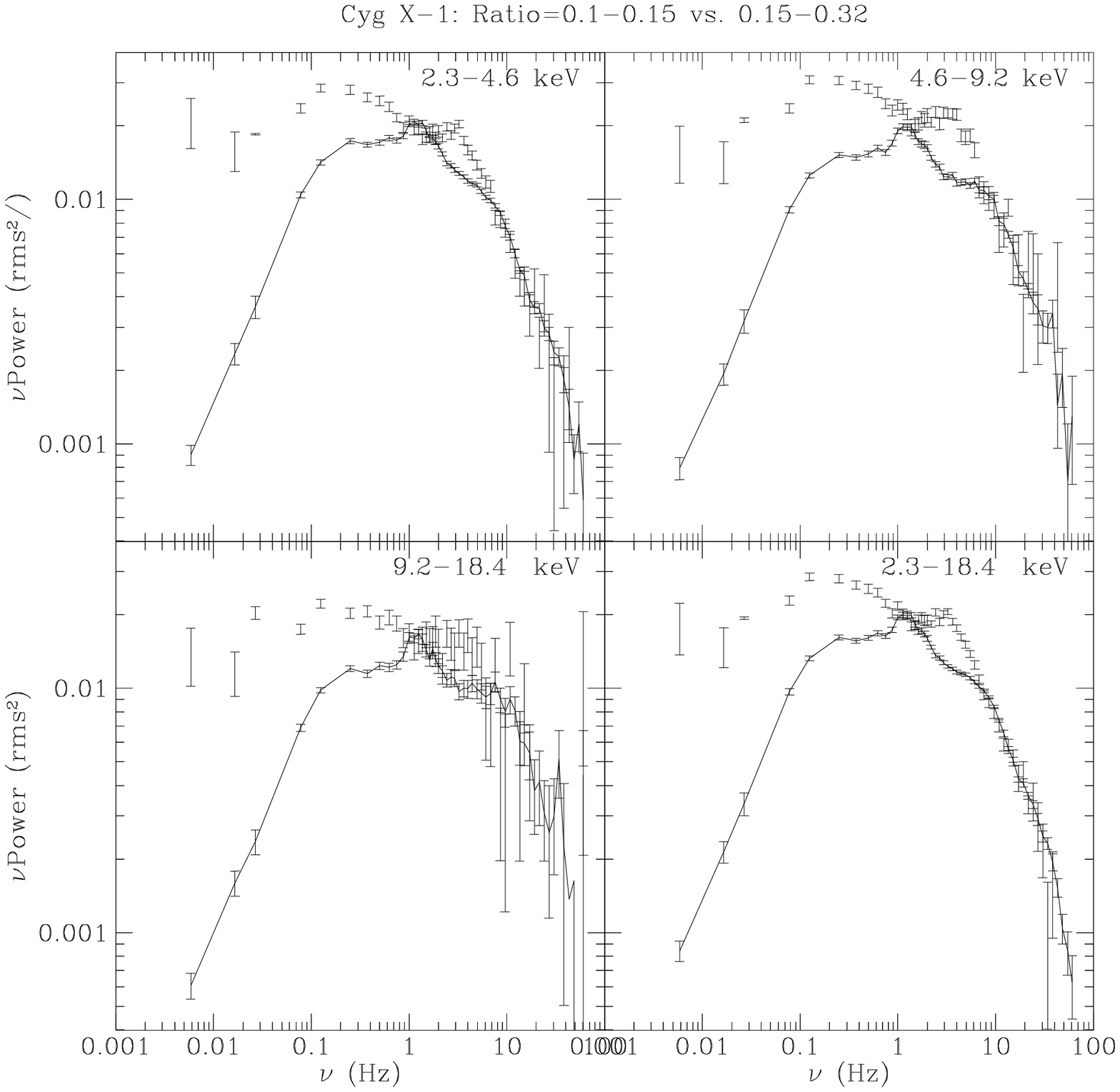 hoffset=-80 voffset=-80}{14.7cm}{21.5cm}
\FigNum{\ref{fig:cygx1splitg}}
\end{figure}


\clearpage
\pagestyle{empty}
\begin{figure}[htb]
\PSbox{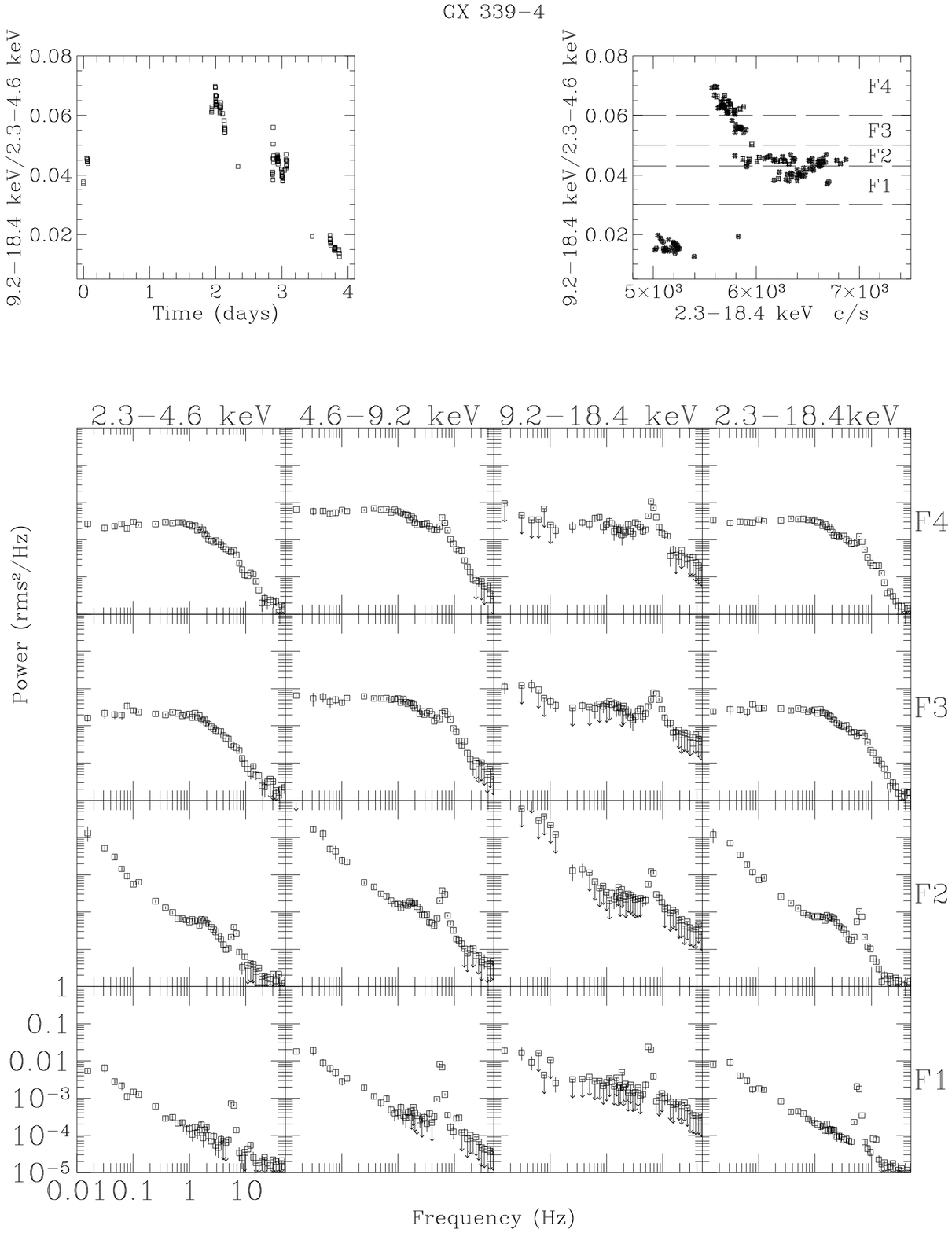 hoffset=-80 voffset=-80}{14.7cm}{21.5cm}
\FigNum{\ref{fig:gx339qpohidpds}}
\end{figure}


\clearpage
\pagestyle{empty}
\begin{figure}[htb]
\PSbox{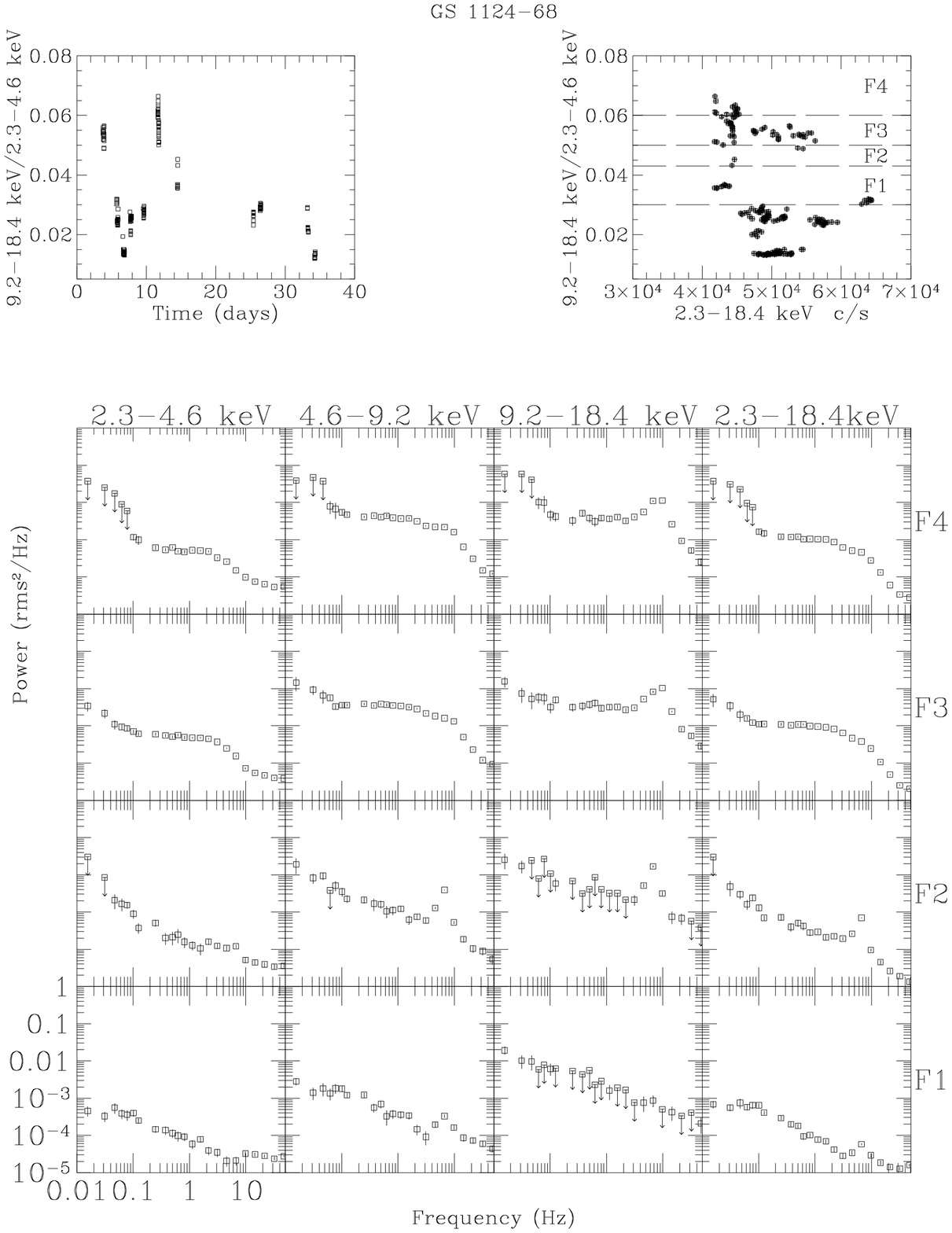 hoffset=-80 voffset=-80}{14.7cm}{21.5cm}
\FigNum{\ref{fig:gs1124qpohidpds}}
\end{figure}


\clearpage
\pagestyle{empty}
\begin{figure}[htb]
\PSbox{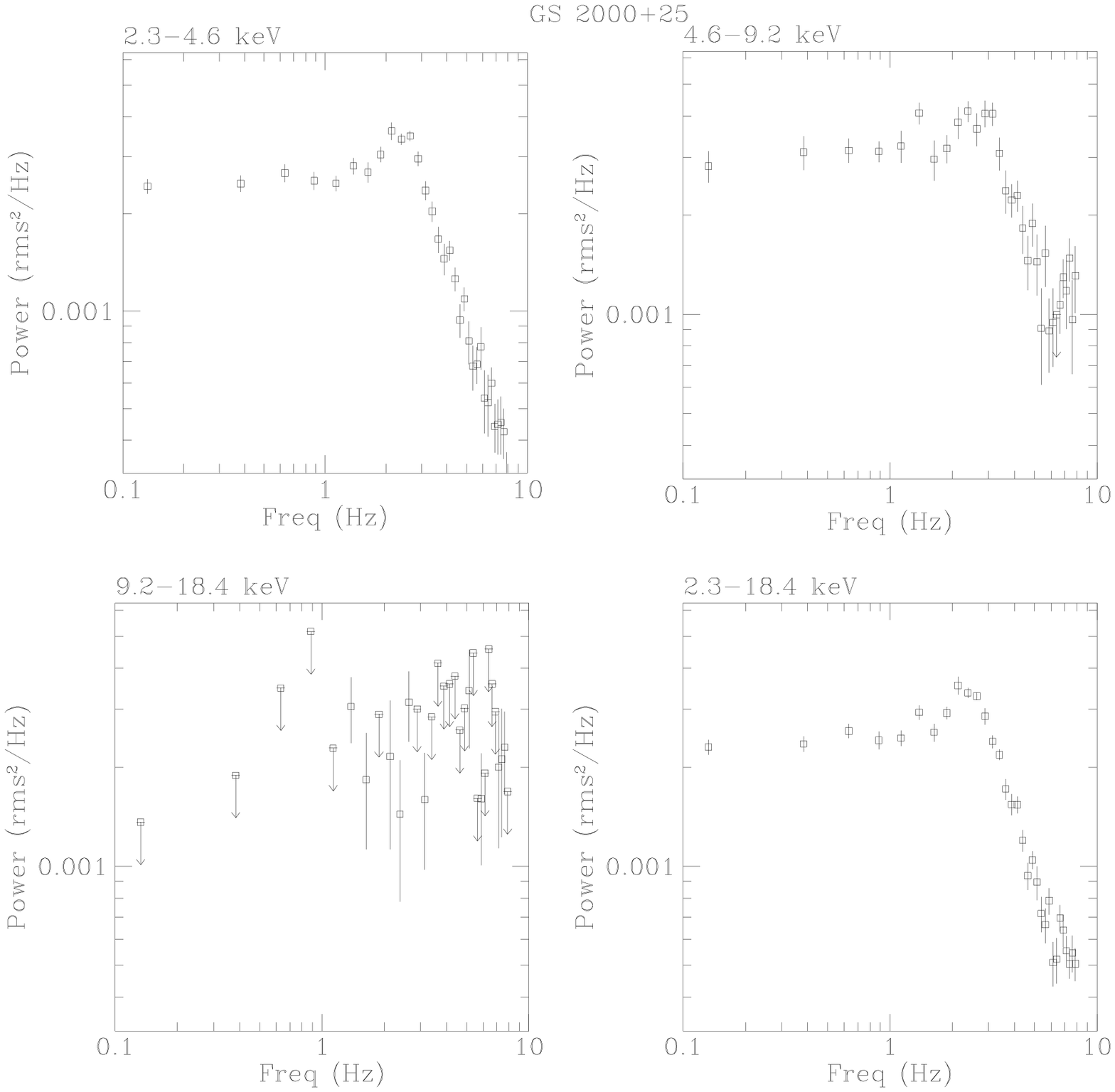 hoffset=-80 voffset=-80}{14.7cm}{21.5cm}
\FigNum{\ref{fig:g2000qpopds}}
\end{figure}


\clearpage
\pagestyle{empty}
\begin{figure}[htb]
\PSbox{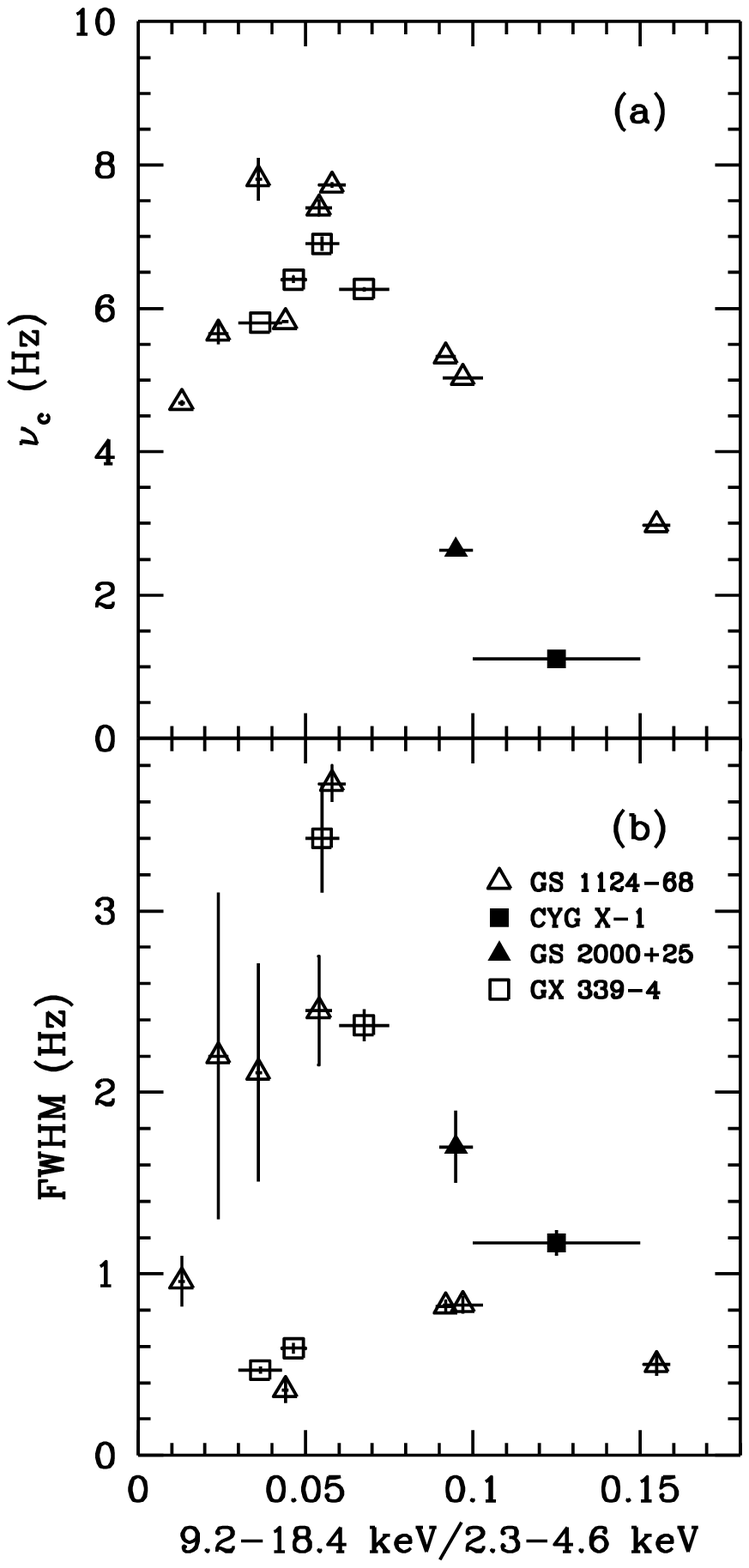 hoffset=-80 voffset=-80}{14.7cm}{21.5cm}
\FigNum{\ref{fig:vhsqpo}}
\end{figure}


\clearpage
\pagestyle{empty}
\begin{figure}[htb]
\PSbox{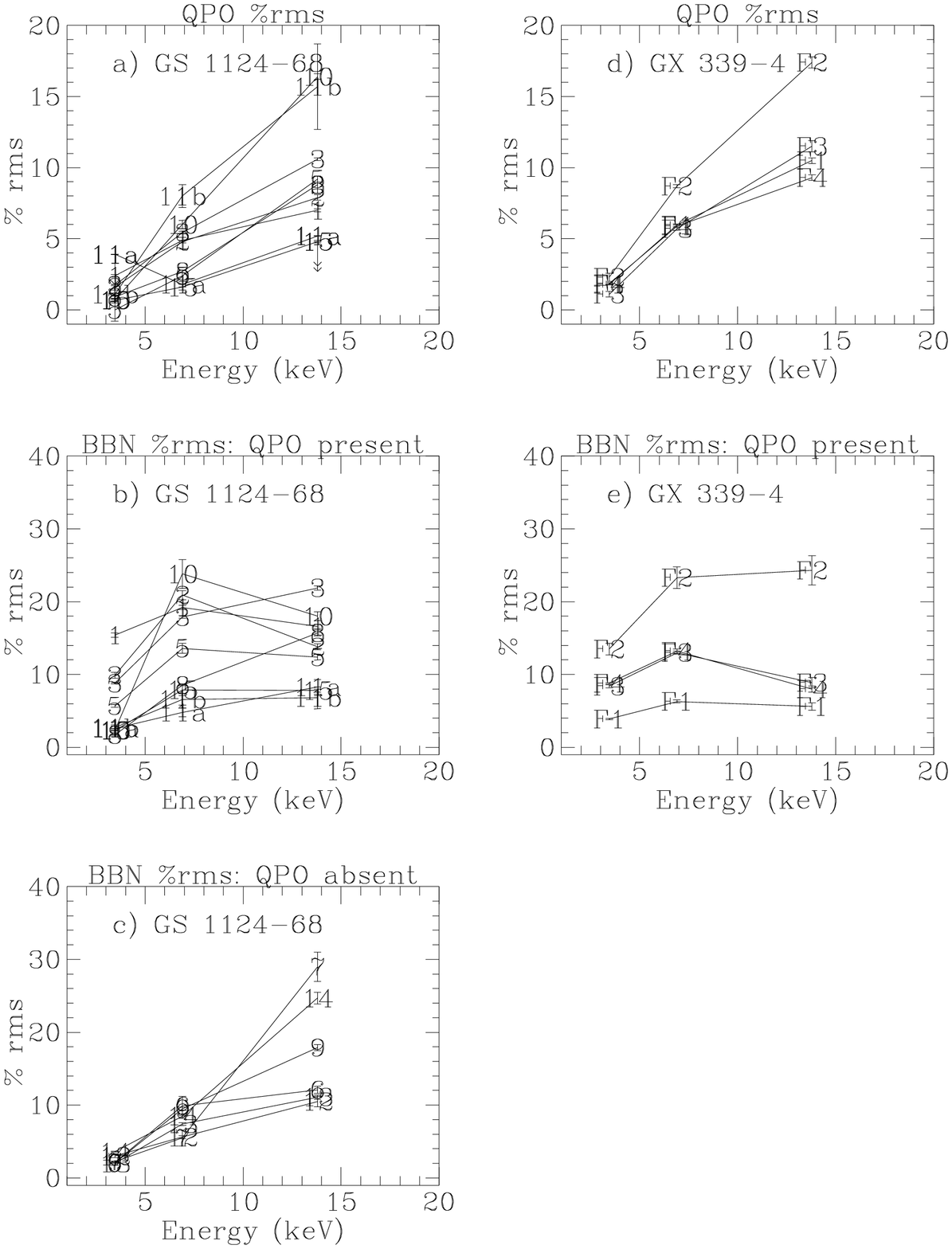  hoffset=-80 voffset=-80}{14.7cm}{21.5cm}
\FigNum{\ref{fig:vhsbbqporms}}
\end{figure}


\clearpage
\pagestyle{empty}
\begin{figure}[htb]
\PSbox{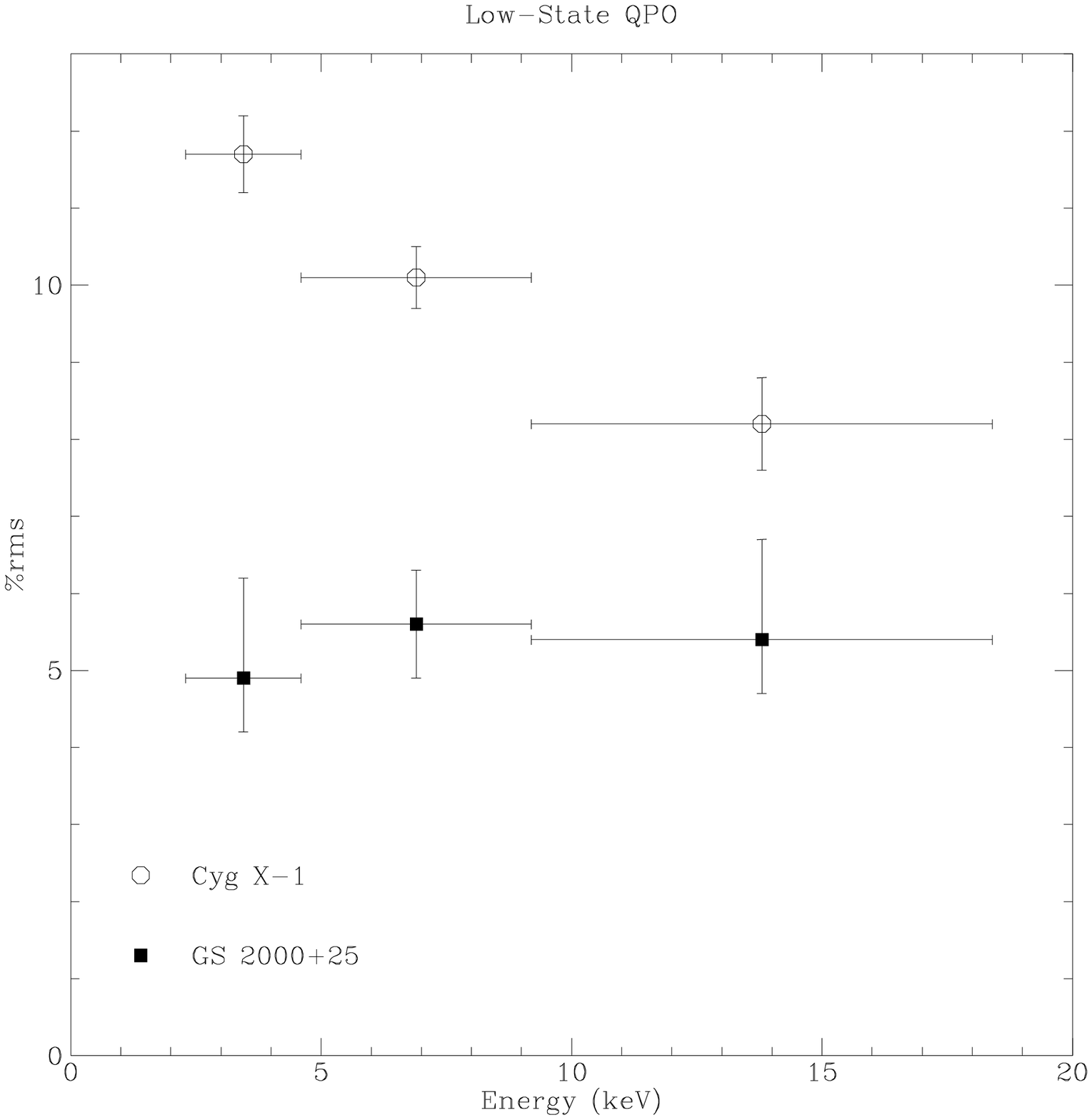 hoffset=-80 voffset=-80}{14.7cm}{21.5cm}
\FigNum{\ref{fig:lsqpo}}
\end{figure}


\clearpage
\pagestyle{empty}
\begin{figure}[htb]
\PSbox{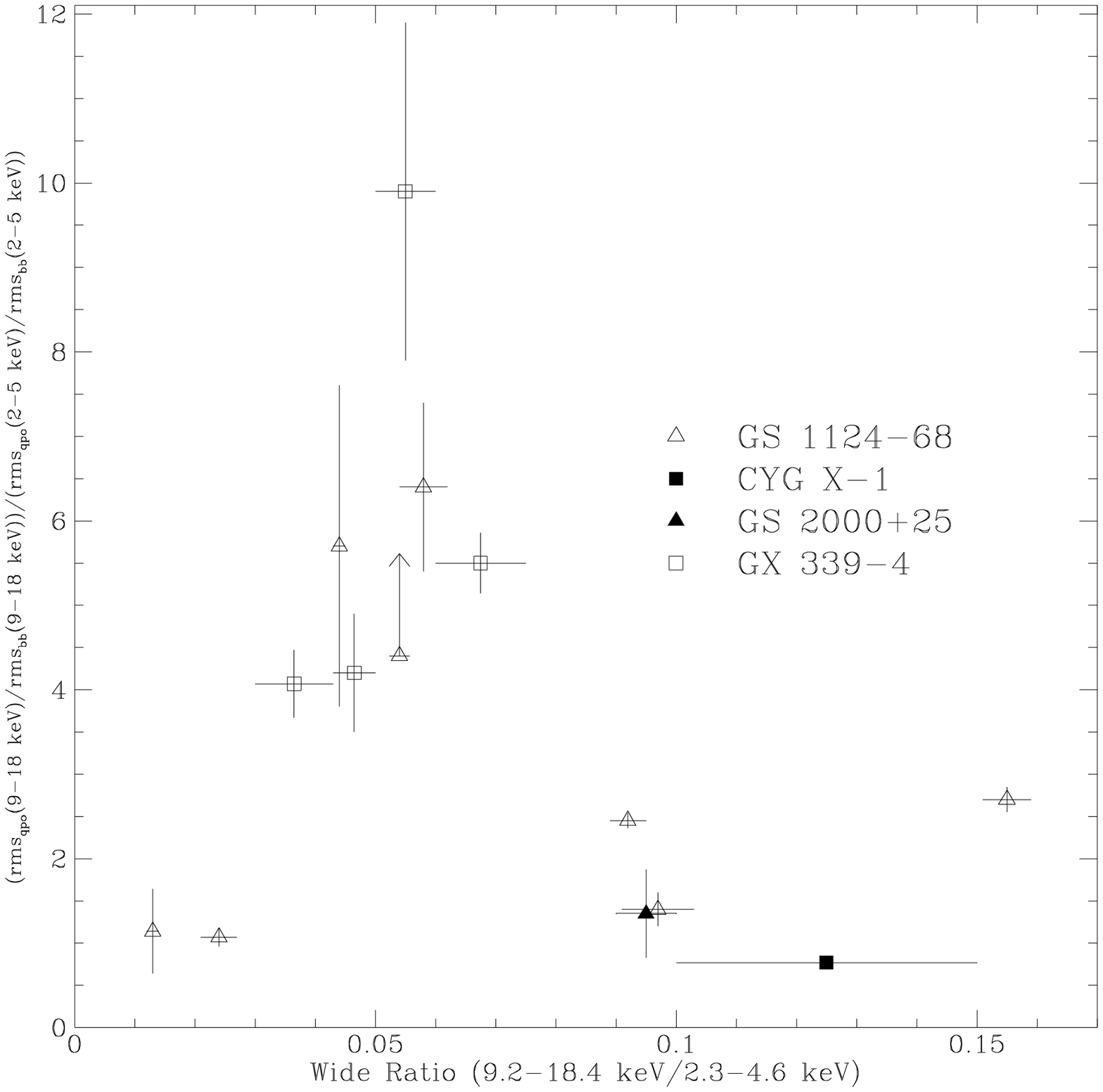  hoffset=-80 voffset=-80}{14.7cm}{21.5cm}
\FigNum{\ref{fig:rrvswr}}
\end{figure}


\clearpage
\pagestyle{empty}
\begin{figure}[htb]
\PSbox{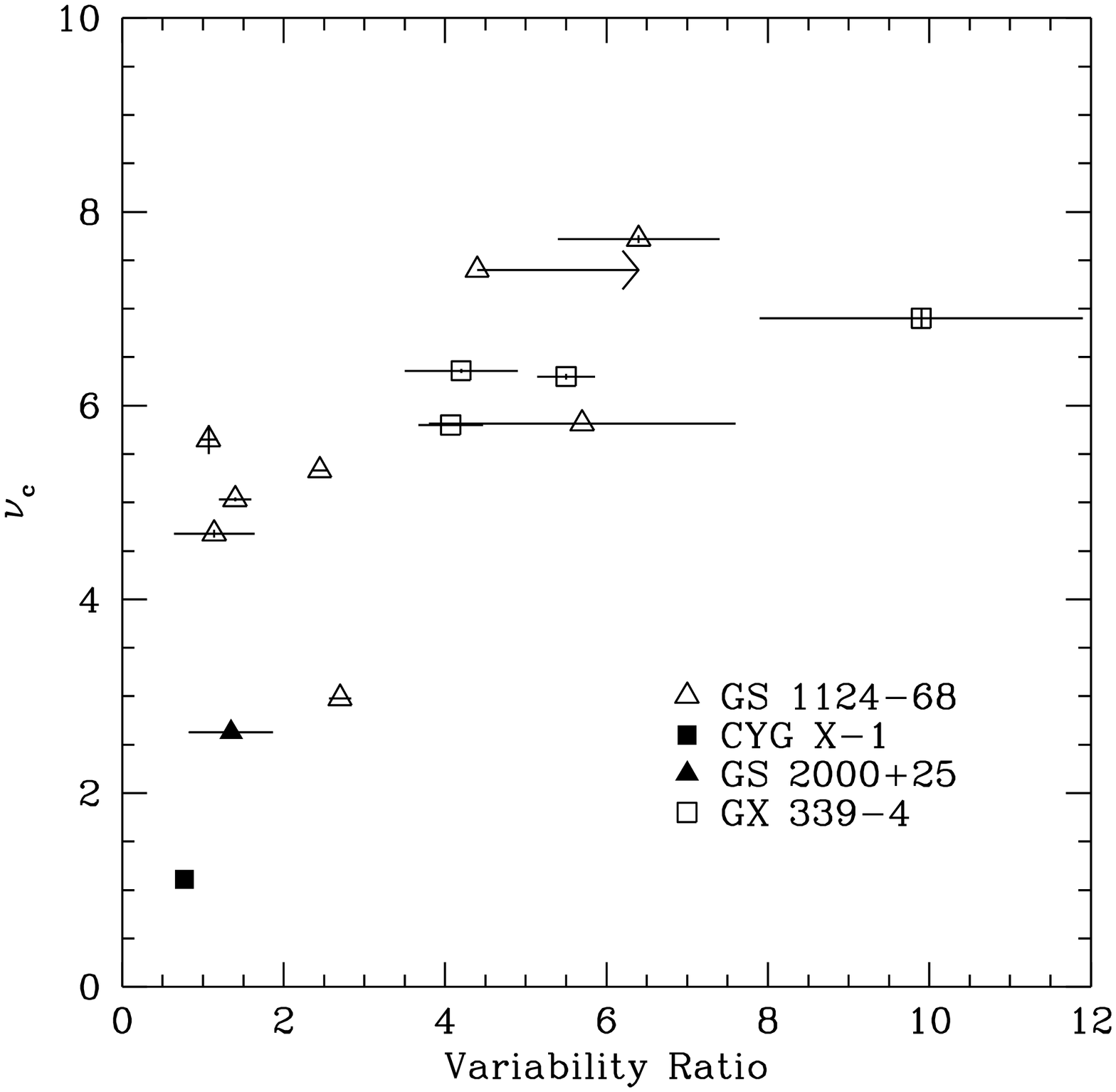  hoffset=-80 voffset=-80}{14.7cm}{21.5cm}
\FigNum{\ref{fig:rrvsvc}}
\end{figure}


\clearpage
\pagestyle{empty}
\begin{figure}[htb]
\PSbox{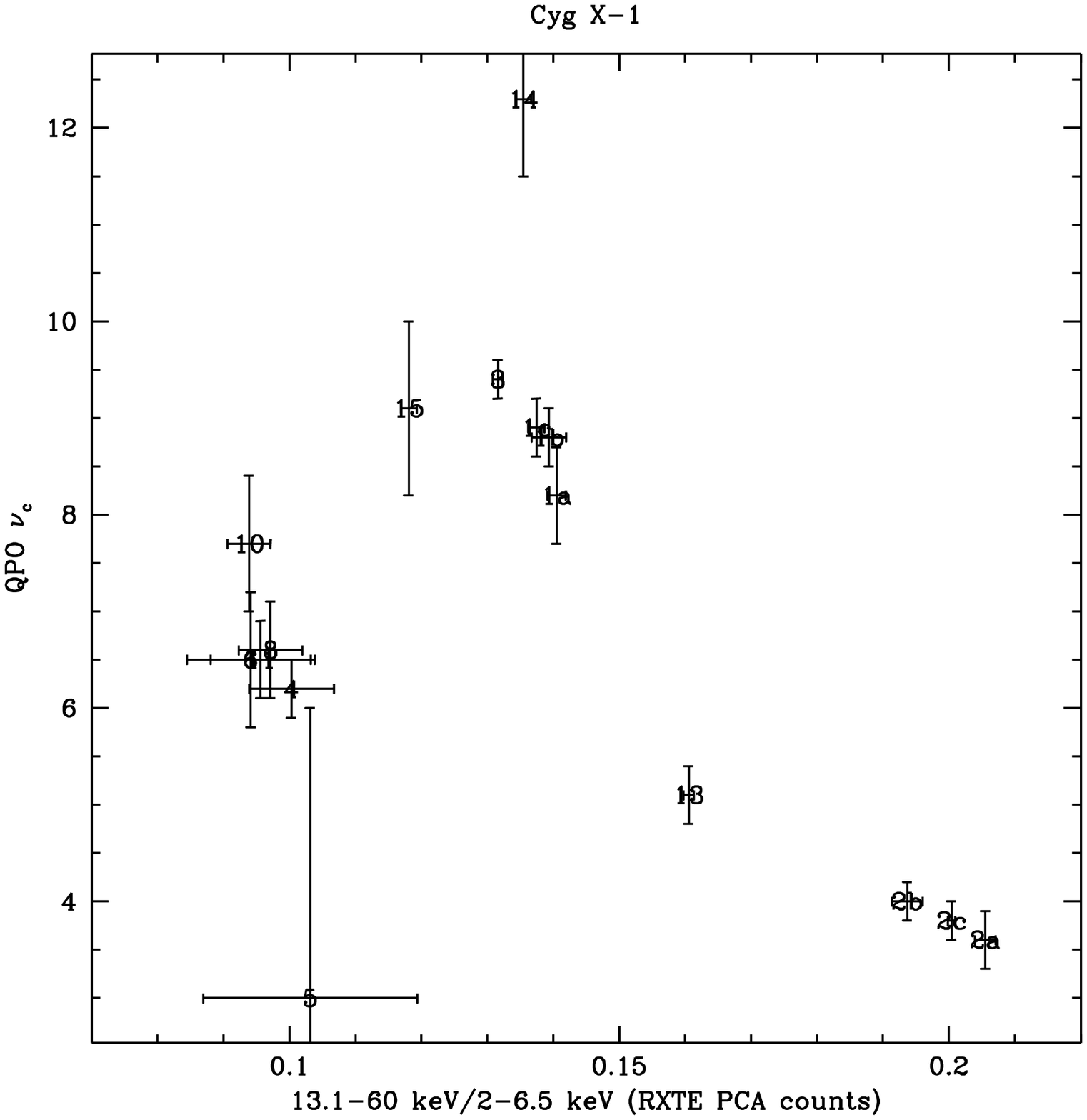  hoffset=-80 voffset=-80}{14.7cm}{21.5cm}
\FigNum{\ref{fig:cui}}
\end{figure}

\end{document}